\documentstyle[12pt,epsfig,axodraw]{article}
\newcommand{\pbarslash}{\not{\!\overline{P}}}

\newcommand{\pnslash}{\not{\!P_N}}
\newcommand{\snslash}{\not{\!S_N}}

\newcommand{\be}{\begin{equation}}
\newcommand{\ee}{\end{equation}}
\newcommand{\ba}{\begin{eqnarray}}
\newcommand{\ea}{\end{eqnarray}}

\newcommand{\nsigma}{\mbox{\boldmath $\sigma$}}

\newcommand{\nl}{{\bf      l}}
\newcommand{\nn}{{\bf      n}}

\newcommand{\nk}{{\bf      k}}
\newcommand{\np}{{\bf      p}}       
\newcommand{\nq}{{\bf      q}}
         
\newcommand{\ns}{{\bf      s}}

\newcommand{\nP}{{\bf      P}}

\begin{document}
\begin{titlepage}
\mbox{} 
\vspace*{2.5\fill} 
{\Large\bf 
\begin{center}
%
Kinematical relativistic effects in $(\vec{e},e'\vec{N})$ reactions
%
\end{center}
} 
\vspace{1\fill} 
\begin{center}
{\large 
M.C. Mart\'{\i}nez$^{1}$,
J.A. Caballero$^{1}$,
T.W. Donnelly$^{2}$
}
\end{center}
\begin{small}
\begin{center}
$^{1}${\sl 
Departamento de F\'\i sica At\'omica, Molecular y Nuclear \\ 
Universidad de Sevilla, Apdo. 1065, E-41080 Sevilla, SPAIN 
}\\[2mm]
$^{2}${\sl 
Center for Theoretical Physics, Laboratory for Nuclear Science 
and Department of Physics\\
Massachusetts Institute of Technology,
Cambridge, MA 02139, USA 
}
\end{center}
\end{small}

\kern 1. cm \hrule \kern 3mm 

\begin{small}
\noindent
{\bf Abstract} 
\vspace{3mm} 

Kinematical relativistic effects are analyzed within the plane-wave
impulse approximation for outgoing nucleon polarized responses in 
coincidence electron scattering. Following recent approaches for
semi-relativistic reductions of the electromagnetic current operator, here an exact treatment
of the problem for the transferred energy and momentum is
presented. Rather than employing existing semi-relativistic expressions for the
single-nucleon matrix elements, in this work the response functions
are expanded directly. Results are compared for different kinematical situations with the
fully-relativistic PWIA calculation as well as with previous expansions.
Off- and `on-shell' prescriptions are also studied in detail, and
a systematic analysis of the various kinematical variables involved in the process is 
presented.

\kern 2mm 

\end{small}

\kern 2mm \hrule \kern 1cm
\noindent MIT/CTP\#3214
\end{titlepage}


\section{Introduction}


The starting point for many analyses
of coincidence electron scattering reactions invokes the plane-wave 
impulse approximation (PWIA). Here a single nucleon in the
nucleus absorbs the total energy and momentum transferred by the virtual photon
(Born approximation), and is subsequently ejected without interacting with the
residual nucleus. Obviously, PWIA is an oversimplified description of the
reaction mechanism. Other ingredients such as final-state interactions (FSI),
Coulomb distortion of the electrons, two-body correlations, ... can be
important when making detailed comparisons
with experimental data. However, the
great advantage of PWIA is that it allows one to simplify and clarify the essential physics issues underlying the problem. Indeed, some observables can be shown to be rather insensitive to FSI and
other distortion effects and effects beyond the impulse approximation are minimal for well-chosen kinematical conditions, and thus PWIA calculations may sometimes be adequate.

A large fraction of past theoretical work on $(e,e'N)$ reactions was
carried out on the basis of non-relativistic calculations. Within this scheme,
the bound and ejected nucleons are described by 
non-relativistic wave functions which are solutions of the Schr\"odinger
equation with phenomenological potentials.
The current operator is also described by a non-relativistic
expression derived directly from a Pauli reduction that begins with the relativistic
current operator and free Dirac spinors. Such standard non-relativistic
reductions have usually been based on expansions in
powers of $p/M_N$, $q/M_N$ and $\omega/M_N$, where $p$ is the missing
momentum, $q$ and $\omega$ are the momentum transfer and energy transfer,
respectively, and $M_N$ is the nucleon mass. This approach constitutes
the basis for the standard distorted-wave impulse approximation (DWIA) that has
been widely used to describe $(e,e'N)$ experiments performed at intermediate energies~\cite{Bof96,Kel96,Frul85}. 

In the last decade, given new higher-energy facilities, some experiments performed have involved momenta and
energies that are high enough to invalidate the
non-relativistic expansions assumed in DWIA. Thus, a consistent description of these
processes requires one to incorporate relativistic degrees of freedom wherever possible~\cite{Pick87,Pick89}. For example,
nuclear responses and cross sections have been investigated recently using the 
relativistic mean field approach~\cite{Udi93,Udi95,Udi96,Udi99,Udi01}. 
This constitutes the basis of the relativistic
distorted-wave impulse approximation (RDWIA), where bound and scattered wave functions
are described as Dirac solutions with scalar and vector potentials, and the 
relativistic free nucleon current operator is assumed. So far, RDWIA calculations
have clearly improved the comparison with experimental 
data over the previous non-relativistic approaches~\cite{Udi93,Udi95,Udi96,Udi99,Udi01,Jin}.

Following the discussion presented in previous work~\cite{Udi99,Udi01,Cris1}, 
relativistic effects can be cast into two
general categories: kinematical and dynamical effects. Dynamical relativistic effects
come from the difference
between the relativistic and non-relativistic nucleon wave functions. For instance, one may distinguish effects associated with the Darwin term, that 
mainly affects the determination of spectroscopic factors at low 
missing momenta~\cite{Udi93,Udi95}, and effects due to the dynamical enhancement of 
the lower components in the relativistic wave functions. These latter effects have been
studied in detail in recent work within the plane-wave 
limit~\cite{Cris1,Cab98a,Cab98b} and
including also FSI between the ejected nucleon and the
residual nucleus~\cite{Udi99,Udi01}. In both cases the effects introduced by the presence of
negative-energy components in the nucleon wave functions (which are generally significant mainly at high missing momentum) have been also proven to be very important for specific
observables even at low/moderate values of the missing momentum $p$. In particular, within
the relativistic plane-wave impulse approximation (RPWIA), the analysis of dynamical 
relativistic
effects is considerably simplified and analytic expressions that explicitly incorporate the contributions arising from the negative-energy components in the bound
nucleon wave function can be obtained.
This subject was thoroughly developed in~\cite{Cab98a}
(see also~\cite{Gardner94}) for
the case of unpolarized $A(e,e'N)B$ reactions, and has recently been extended to the
case of recoil nucleon polarized processes~\cite{Cris1}. The analysis of the
unpolarized $(e,e'p)$ reaction including FSI was discussed in~\cite{Udi99,Udi01} and
its extension to $A(\vec{e},e'\vec{N})$ will be presented in a forthcoming 
publication~\cite{Cris3}.

Our main interest in this paper is focused on the analysis of the kinematical relativistic
effects. These are directly connected to the structure of the four-component current operator
compared with the non-relativistic (two-component) expressions. The usual
procedure to build the electromagnetic nuclear current operators 
starts by employing the on-shell single-nucleon currents. In~\cite{Ama96}
the exact on-shell operators for use between two-component spin spinors were developed.
For applications to nuclear physics where nucleons are off-shell, it is necessary
to make further approximations. First, the on-shell expressions for the operators
can be inserted between nuclear wave functions, which are not on-shell plane waves
but are the off-shell single-particle wave functions describing the
nucleons interacting in nuclei. 
Second, the standard procedure has been to expand
the expressions for the electromagnetic current in dimensionless momenta,
retaining only the leading-order terms in all energies or momenta over
nucleon mass. This gives rise to the standard
non-relativistic limit, yielding the simple expressions obtained for the electromagnetic
currents that have been commonly used for many years in DWIA analyses. Unfortunately, these approximations
are not adequate in present high-energy experiments, as $q$ and $\omega$ can be comparable to $M_N$. 

In recent work~\cite{Ama96,Ama98} new expressions for the current
operators were deduced treating the transferred energy and
momentum exactly while expanding only in missing momentum over nucleon
mass. The results to date show that these
`relativized' current operators retain important relativistic aspects not taken
into account in the traditional non-relativistic reductions. While we
frequently continue to call this approach a non-relativistic expansion, it
might be more appropriate to call it ``semi-relativistic'' in that
part of the (kinematical) relativistic behaviour is being taken into
account, leaving only a variable such as $\chi\equiv \frac{p}{M_N}\sin\theta$, where $\theta$ 
is the angle formed by the missing momentum $\np$ and the momentum
transfer $\nq$ in which to expand. Clearly for some circumstances $\chi$ is indeed small, whereas
both $q/M_N$ and $\omega/M_N$ are not, and it might be expected that
this approach will go a long way towards incorporating at least some
aspects of relativity in the analysis.

In this work we build on these ideas and deduce new relativized expressions for the 
polarized (and unpolarized) responses that enter in the analysis of 
$A(\vec{e},e'\vec{N})B$ reactions within PWIA. In contrast to the previous 
analyses~\cite{Ama96,Ama98} where the semi-relativistic expansion was performed at the level of the
single-particle current matrix elements, here we directly expand the single-nucleon
responses. 

It is important to point out that, even treating the problem of the transferred
energy and momentum exactly, there are ambiguities in doing the semi-relativistic expansions.
Different choices of the variables in which to expand can be made, and moreover,
momentum-energy conservation relations imposed on the different kinematical variables should
be treated with caution as soon as a semi-relativistic expansion is considered in one of 
the variables. Several options have been explored in this work and the results obtained
are compared with the fully-relativistic PWIA calculation as well as with previous
expansions. Our main aim is to establish how
precise the semi-relativistic reduction is, and under which conditions and/or for which
observables it does or does not work.

Despite such ambiguities, in this paper most of the analysis is based on the choice of the expansion 
variable $\chi$ as in fact there exist reasons to choose it over some of
the alternatives. First, this is the variable that enters in
a natural way in the fully-relativistic PWIA expressions for the unpolarized and polarized
single-nucleon responses. Second, its behaviour for small values of the missing momentum,
independent of the kinematics selected, allows one to be confident about the
accuracy of the semi-relativistic
expansion. Indeed, using $\chi$ we shall see that the various responses can be grouped in a natural way 
in different classes according to their
leading order term $\chi^n$. Class ``0'' responses correspond to those whose 
leading order is given by $\chi^0$, class ``1'' to leading order $\chi$, class ``2'' to $\chi^2$, and so on. 
We study how ``safe'' we expect the semi-relativistic reductions in
$\chi$ to be for the three classes of response. Third, the kinematical
effects of relativity that form the basis of the present work are not
the end of the story: in work being done in parallel~\cite{Cris1}
explorations are being made of dynamical relativistic effects and the same
organization into classes of response appears in a natural way when
$\chi$ forms the basis for the expansion strategy.

Another issue directly connected with the uncertainty introduced in the analysis of the
response functions is the off-shell character of the nucleons involved in $(e,e'N)$
processes, particularly for the initial-state bound nucleon. At present, the usual
way to deal with this problem is to use different {\it ad hoc} recipes (prescriptions)
for the current operator that may include some aspects of current conservation and Lorentz
covariance. In this way we may obtain some insight into the importance of the off-shell
effects, although it should always be kept in mind that the off-shell
problem is not really fully under control. Here we proceed in the
commonly accepted way and, lacking a fundamental approach to follow,
continue to invoke the familiar off-shell prescriptions. 

Finally, we also explore the so-called
`on-shell' prescription where the bound nucleon is forced to be on-shell. This leads,
from energy-momentum conservation, to an unphysical negative excitation energy. However,
the important advantage of the `on-shell' prescription is that it does
not present either Gordon or gauge 
ambiguities --- the current is naturally conserved.
An estimate of the uncertainty
introduced by the semi-relativistic reductions is provided by comparing the results
obtained using the new `relativized' single-nucleon responses with the `on-shell'
and various off-shell fully-relativistic PWIA calculations.

The present study has been undertaken in a wider context where various
other aspects of the problem of treating relativistic effects in
$(e,e'N)$ reactions provide the focus. Any attempt to put all of the
facets of the overall study together would be far too unwieldy and
accordingly the various components of our work are being presented
separately. It is, however, important to understand how they interact
with one another:
\begin{itemize}
\item The present work addresses only the issue of {\bf kinematical}
  relativistic effects for responses and polarization asymmetries
  strictly within the context of PWIA. Both off-shell and `on-shell'
  approaches are examined.
\item In an accompanying paper~\cite{Cris1} {\bf dynamical} relativistic effects --- those
  stemming from having non-trivial relativistic content in the nuclear
  wave functions --- are also studied, specifically for relativistic mean field
  bound-state wave functions, but with plane-wave final ejected
  nucleon wave functions (RPWIA).
\item In work in progress~\cite{Cris3} relativistic kinematical and dynamical
  effects are being addressed with all of the above ingredients plus
  {\bf relativistic FSI} effects, i.e., the RDWIA.
\end{itemize}

The paper is organized as follows: in Section~2 we present the basic formalism
needed to describe $A(\vec{e},e'\vec{N})B$ reactions, paying special attention to the analysis of the
kinematics involved in coincidence electron scattering reactions. We discuss various
options for selecting the independent kinematical variables that completely define the $(e,e'N)$ process. The relations held by these variables coming from energy-momentum
conservation are also discussed. In Section~3 we focus on the PWIA. The fully-relativistic
polarized single-nucleon responses corresponding to the 
different off-shell prescriptions are derived, and we also introduce the `on-shell'
approach. In this section we also derive and 
discuss the semi-relativistic reductions of the responses.
The various off-shell PWIA responses compared with the
semi-relativistic and `on-shell' results evaluated for different kinematics are presented
in Section~4. Finally in Section~5 we present a summary and our conclusions.


\section{Formalism for $A(\vec{e},e'\vec{N})B$ reactions}



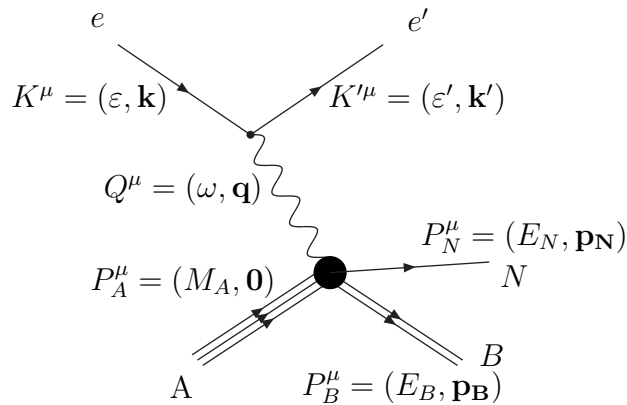
\begin{figure}
\begin{center}
\begin{picture}(300,200)(0,0)
\Text(80,177)[l]{$e$}
\ArrowLine(90,167)(140,133)
\Text(50,147)[l]{$K^{\mu}=(\varepsilon,\mathbf{k})$}
\Text(200,177)[l]{$e'$}
\ArrowLine(140,133)(190,167)
\Text(170,147)[l]{$K'^{\mu}=(\varepsilon',\mathbf{k'})$} 
\Vertex(140,133){1.5}
\Photon(140,133)(170,81){4}{4.5}
\Text(85,113)[l]{$Q^{\mu}=(\omega,\mathbf{q})$}
\ArrowLine(170,81)(220,48)
\ArrowLine(168,79)(218,46)
\Text(227,50)[l]{$B$}
\ArrowLine(120,48)(170,81)
\ArrowLine(118,50)(168,83)
\ArrowLine(122,46)(172,79)
\Text(110,38)[l]{A}
\Text(80,78)[l]{$P^{\mu}_A=(M_A,\mathbf{0})$}
\GCirc(170,81){6}{0}
\ArrowLine(170,81)(230,85)
\Text(235,80)[l]{$N$}
\Text(205,95)[l]{$P^{\mu}_{N}=(E_{N},\mathbf{p_{N}})$}
\Text(160,38)[l]{$P^{\mu}_{B}=(E_{B},\mathbf{p_{B}})$}
\end{picture}
\end{center}
\caption{\label{figure1}Feynman diagram for the $A(e,e'N)B$ process within 
the Born approximation.}
\end{figure}

In this section we briefly summarize the basic formalism involved in the description of
$(e,e'N)$ reactions (see also~\cite{RaDo89,Donnelly,DoIn99}). The Feynman diagram within the Born
approximation (one virtual photon exchange)
is depicted in Fig.~\ref{figure1} and defines our conventions on
energies and momenta. Apart from assuming the Born approximation, in what follows
no assumption is 
made about the reaction mechanism; we simply introduce the kinematical variables
that completely specify the process and use energy-momentum conservation to 
inter-relate the various energies and momenta:
\ba
\nq &=& \nk-\nk' = \np_N+\np_B \label{eq1s2}
\\
\omega &=& \varepsilon-\varepsilon' = E_B+E_N-M_A \, ,
\label{eq2s2}
\ea
where the target is assumed to be at rest in the laboratory frame.
The missing momentum $\np$ is defined as
\be
\np\equiv -\np_B=\np_N-\nq \, ,
\label{eq3s2}
\ee
where $p=|\np|$ characterizes the split in momentum flow between the detected
nucleon and the unobserved daughter nucleus. The corresponding
split in energy can be characterized by the 
excitation energy of the residual nucleus
\be
{\cal{E}} \equiv E_B-E_B^0 \geq 0 \, ,
\label{eq4s2}
\ee
where $E_B=\sqrt{p_B^2+M_B^2}$ and $E_B^0=\sqrt{p_B^2+M_B^{0^2}}$. Here $M_B$ includes the
internal excitation energy of the residual system, while $M_B^0$ is its rest mass in its ground
state.

Now let us briefly discuss the set of six independent kinematical
variables that completely describe the cross section for the $(e,e'N)$ process. 
First, explicit dependences on the electron scattering angle
$\theta_e$ (through the general Rosenbluth factors) and the azimuthal angle $\phi$ 
can be isolated (see~\cite{RaDo89} for details). 
Once the azimuthal angle $\phi$ and the electron scattering angle $\theta_e$ are
fixed, the $(e,e'N)$ cross section is totally
specified by four kinematical variables, for instance $\{q,\omega,{\cal{E}},p\}$. 
The dependences of the responses on these 
so-called `dynamical' variables 
involve detailed aspects of the nuclear current matrix elements, in contrast to the 
dependences on $\theta_e$ and $\phi$ which are effectively geometric. Obviously, various alternative
sets of four `dynamical' variables are possible when studying specific
$(e,e'N)$ reactions. In what follows we present a brief discussion of some
alternatives that have advantages for specific choices of kinematics.

\begin{figure}
{\par\centering \resizebox*{0.7\textwidth}{0.82\textheight}{\includegraphics{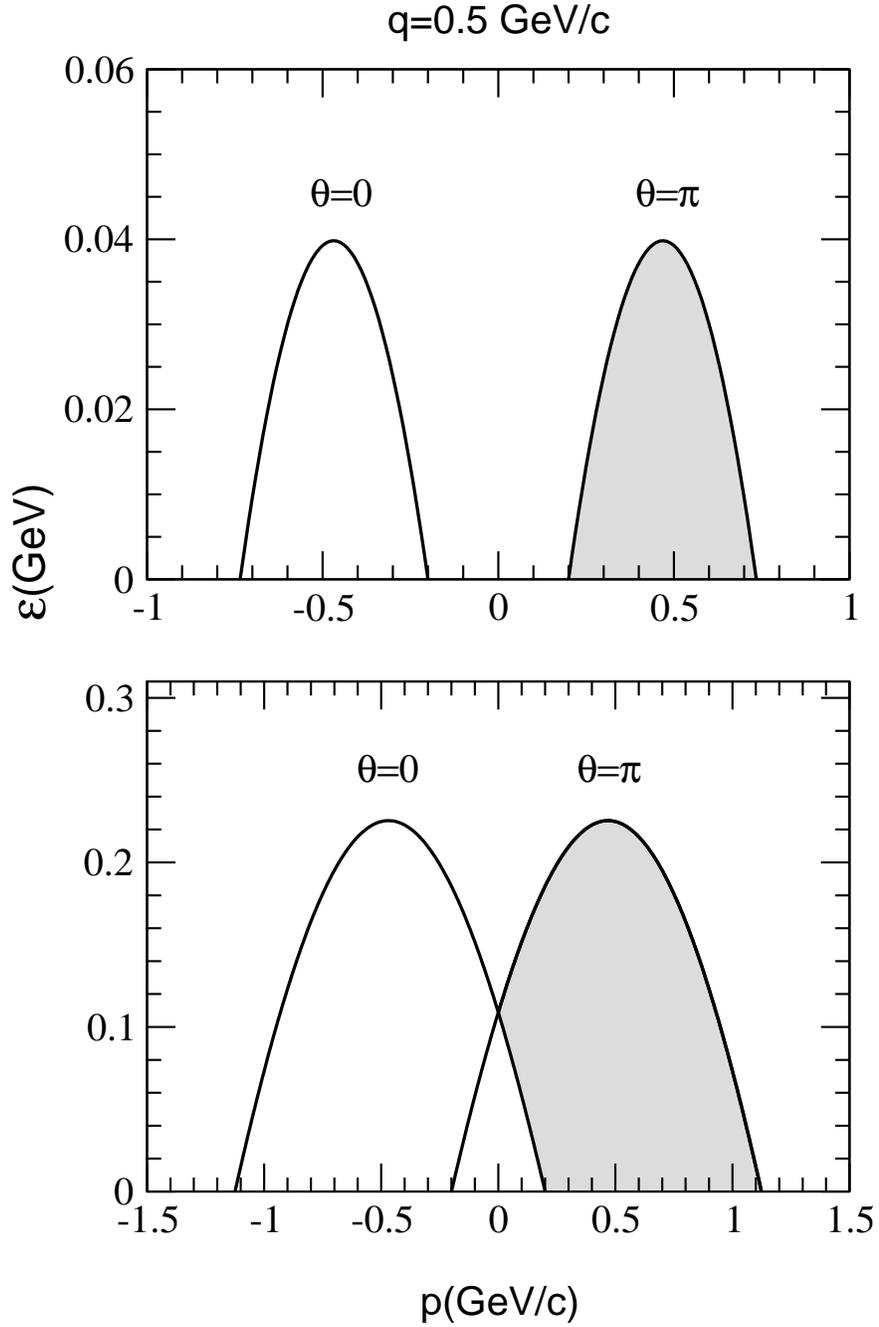}} \par}
\caption{\label{figure2}Excitation energy as a function of the missing momentum $p$ for the process \protect\( ^{16}O(e,e'p)^{15}N\protect \). In the top panel the momentum transfer $q$
is fixed to \protect\( q=0.5\protect \) GeV/c and $\omega =55.0$ MeV $< \omega_{QEP}$, while in the bottom panel it is taken to be $\omega =240.6$ MeV $>\omega_{QEP}$.}
\end{figure}

Let us
begin assuming that the momentum and energy transferred in the process are fixed. Then,
the excitation energy $\cal{E}$ in terms of $q$, $\omega$,
$p$ (missing momentum) and the angle $\theta$ (between $\np$ and $\nq$) is given by
\be
{\cal{E}} = M_A+\omega-\sqrt{M_N^2+p^2+q^2+2pq\cos\theta}-
\sqrt{p^2+M_B^{0^2}} \, .
\label{eq9s2}
\ee
Thus there are clear relationships between the sets $\{E_N,\theta_N\}$ ($\theta_N$ being the angle between $\mathbf{p_N}$ and $\mathbf{q}$)  and
$\{p,\theta\}$ and hence $\{{\cal{E}},p\}$. Note that Eq.~(\ref{eq9s2}) yields a curve of
$\cal{E}$ versus $p$ for each choice of $\theta$, where the requirement 
$|\cos\theta|\leq 1$ imposes constraints on the kinematics. Let us recall what domain in the $({\cal{E}},p)$ plane is compatible with the conservation of energy
and momentum. First, we introduce the quasielastic peak value for the energy transfer $\omega_{QEP}$ which is given by
\be
\omega_{QEP}=\sqrt{q^2+M_N^2}+M_B-M_A \, .
\label{eq10s2}
\ee
For $\omega$-values such that $\omega<\omega_{QEP}$ the trajectories $\cos\theta=\pm 1$
for a selected value of the momentum transfer $q$ are plotted in the top panel in Fig.~\ref{figure2}. 
Here we only show
the physical region ${\cal{E}} \geq 0$. All physically allowable values of $\cal{E}$ and
$p$ must lie below the curve $\cos\theta=-1$ and, of course, above ${\cal{E}}=0$. In contrast,
in the case of the curve corresponding to $\cos\theta=+1$, the allowable values of $\cal{E}$ and
$p$ must lie above this curve, and thus no physically values occur for this condition. Therefore, for $\omega <\omega_{QEP}$ the physical region is completely defined
by the $\cos\theta=-1$ curve and ${\cal{E}}=0$ (shadowed region in top panel in Fig.~\ref{figure2}). 
The limit values
of the missing momentum $p$ are denoted by $p_{min}\equiv -y \geq 0$ and $p_{max}\equiv +Y$
(see~\cite{Donnelly} for the explicit expressions).
Note that $\omega<\omega_{QEP}$ then implies that $y<0$. 

The case $\omega>\omega_{QEP}$ is shown in the bottom panel in Fig.~\ref{figure2}. 
Here the $\cos\theta=-1$ curve is similar, except that now $p_{min}$ is negative and so $y$ is positive. As in the previous case
the physically allowable region must lie below the $\cos\theta=-1$ curve and above the
$\cos\theta=+1$ curve. Note however that the latter condition does enter in the quadrant where ${\cal{E}}\geq 0$
and $p\geq 0$, providing a new boundary condition (see shadowed region).

In both cases (positive and negative $y$-regions) the value of the energy transfer $\omega$
is completely specified in terms of $q$ and $y$
\be
\label{eq:w(q,y)}
\omega (q,y)=\sqrt{M_{N}^{2}+(q+y)^{2}}+\sqrt{y^{2}+M^{2}_{B}}-M_{A} \, ,
\ee
where $\omega=\omega_{QEP}$ occurs for $y=0$. Thus, the set of four `dynamical' variables
might be chosen as $\{q,\omega,E_N,\theta_N\}$, $\{q,\omega,{\cal{E}},p\}$  or equivalently (sometimes more conveniently)
$\{q,y,{\cal{E}},p\}$.

Having set up the general kinematics involved in coincidence electron
scattering reactions, next we consider briefly the general form for the
coincidence cross section with the incident electron beam and outgoing nucleon
both polarized. Within the Born approximation, the general cross section in
the laboratory system can be written as
\be
\frac{d\sigma }{d\varepsilon'd\Omega _{e}d\Omega _{N}}=
\frac{2\alpha ^{2}}{Q^{4}}\left(\frac{\varepsilon'}{\varepsilon}\right)
Kf_{rec}^{-1}\eta_{\mu\nu}W^{\mu\nu} \, ,
\label{eq1s3}
\ee
with $K$ a kinematical constant given by $K=p_N M_N M_B/M_A$, $\alpha$ is the fine structure constant,
$\eta_{\mu\nu}$ is the familiar leptonic tensor that can be evaluated
directly using trace techniques~\cite{RaDo89}, and $W^{\mu\nu}$ is the
hadronic tensor containing all of the nuclear structure and dynamics information.
The latter is given directly in terms of the nuclear electromagnetic transition
currents in momentum space.

Using the general properties of the leptonic tensor, it may be shown that the
contraction of the leptonic and hadronic tensors can be decomposed in terms of
leptonic kinematical ``super-Rosenbluth'' factors and response functions. The differential cross section may then be written
\be
\frac{d\sigma}{d\Omega_e d\varepsilon' d\Omega_N}=
K\sigma_{M}f_{rec}^{-1}
\left[v_LR^L+v_TR^T+v_{TL}R^{TL}+v_{TT}R^{TT}+h\left(v_{T'}R^{T'}+v_{TL'}R^{TL'}\right) \right], 
\label{responses}
\ee
where $f_{rec}$ is the usual
recoil factor~\cite{RaDo89} and $\sigma_M$ is the Mott cross section. The kinematic factors
$v_\alpha$ contain all of the dependence on the leptonic vertex aside from overall multiplicative factors (see~\cite{RaDo89} for their explicit expressions in the extreme
relativistic limit (ERL)). The factor $h=\pm 1$ is the incident
electron's helicity. The hadronic current enters via the response functions $R^\alpha$. The
labels $L$ and $T$ refer to projections of the current matrix elements
longitudinal and transverse to the virtual photon direction,
respectively. 

In the case of $A(\vec{e},e'\vec{N})B$ reactions, the hadronic response functions are
usually given by referring the recoil nucleon polarization to the
coordinate system defined by the
axes\footnote{Note that for simplicity in this work we use the notation $l$, $n$,
  $s$, whereas in~\cite{RaDo89} final-state polarizations were labelled with
  primes, $l'$, $n'$, $s'$, to distinguish them from initial-state
  target polarizations for which the primes were absent.}: $\nl$ (parallel to the
momentum  $\np_N$ of the outgoing nucleon), $\nn$ (perpendicular to the plane 
containing $\np_N$ and the momentum transfer $\nq$), and $\ns$  (determined by
$\nn \times \nl$). A total of eighteen response functions enter
in the general analysis of $A(\vec{e},e'\vec{N})B$ reactions
(see~\cite{Pick87,Pick89,Giusti89} for their explicit expressions).
In terms of the polarization asymmetries, the differential cross section can
be expressed in the form
\be
\frac{d\sigma}{d\varepsilon' d\Omega_ed\Omega_N}=
\frac{\sigma_0}{2}\left[1+\nP\cdot\nsigma+h\left(A+\nP'\cdot\nsigma\right)\right]\, ,
\label{eq11s3}
\ee
where $\sigma_0$ is the unpolarized cross section, $A$ denotes the
electron analyzing power, and $\nP$ ($\nP'$) represents the induced
(transferred) polarization. 
A general study of the properties and symmetries of all of these responses and
polarizations can be found in~\cite{Pick87,Pick89}. Here, we
simply note that 
in coplanar kinematics, $\phi=0$,
the only surviving polarization components are $P_n$, $P'_l$ and $P'_s$.


\section{Plane-Wave Impulse Approximation }


The PWIA constitutes the simplest approach to describing
the reaction mechanism for coincidence electron scattering reactions. It has been discussed
in detail in previous work~\cite{Bof96,Kel96,Frul85,Cab93}, and thus here we simply summarize the basic
expressions needed for the discussions to follow. The basic
assumptions in PWIA are the following: i) the electromagnetic current is taken to be a
one-body operator (impulse approximation), ii) the ejected nucleon is a plane wave, 
i.e., the nucleon emerges from the nucleus without interaction with
the residual nuclear system, 
and iii) the nucleon detected in the
coincidence reaction is the one to which the virtual photon is attached. 

The analysis of $(\vec{e},e'\vec{N})$ reactions is further
simplified when FSI are neglected.  The induced polarization $\nP$ and the
analyzing power $A$ are zero
in this case~\cite{Pick87,Pick89}. Thus, only the transferred asymmetry
$\nP'$ survives in the plane-wave limit. Moreover, the normal component
$P'_n$ enters only for out-of-plane kinematics. 
In terms of nuclear responses, from the total of eighteen response functions,
only nine survive within
PWIA. Four, $R^L_0$, $R^T_0$, $R^{TL}_0$ and $R^{TT}_0$, 
represent the unpolarized responses and
the five remaining, $R^{T'}_l$, $R^{T'}_s$, $R^{TL'}_l$, $R^{TL'}_s$ and $R^{TL'}_n$, 
depend explicitly
on the recoil nucleon polarization and only enter when the 
electron beam is also polarized. 

An important simplification of PWIA is that the cross section factorizes into two basic terms, the
electron-nucleon cross section (affected by off-shell uncertainties) and the spectral
function or, in the case where the daughter state is a discrete one (as in this work), the momentum distribution that provides the probability of finding a nucleon
in the nucleus with given energy and momentum. 
Within PWIA the hadronic tensor is then given in the form
\be
W^{\mu\nu}={\cal W}^{\mu\nu}(\np,\nq)N(p) \, ,
\label{PWIA}
\ee
where $N(p)$ represents the momentum distribution of a (non-relativistic) bound orbital, and
${\cal W}^{\mu\nu}$ is the tensor for elastic scattering on a free nucleon whose explicit
expression can be evaluated directly using trace techniques~\cite{Cab98a,Cab93}.
It is crucial to point out that the factorization result in 
Eq.~(\ref{PWIA}) is only strictly valid if the dynamical enhancement of the negative-energy
components in the bound wave function is neglected, i.e., if the bound wave function is 
expanded in terms of free positive-energy Dirac spinors $u$ alone. The contribution
of the negative-energy components gives rise to the relativistic plane-wave impulse
approximation (RPWIA) and destroys the factorization property 
(see~\cite{Cris1,Cab98a,Cab98b}). 

The differential cross section can be written in PWIA as
\be
\frac{d\sigma}{d\Omega_ed\varepsilon'd\Omega_N} =
K f_{rec}^{-1}\sigma^{eN}N(p) \, ,
\label{CSPWIA}
\ee
with $\sigma^{eN}$ the electron-nucleon cross section that can be decomposed into
single-nucleon response functions according to
\be
\sigma^{eN}=\frac{2\alpha^2}{Q^4}\frac{\varepsilon'}{\varepsilon}\eta_{\mu\nu}
{\cal W}^{\mu\nu}(\np,\nq) =
\sigma_M\left[\sum_{\alpha=L,T,TL,TT}v_\alpha{\cal R}^\alpha +h
             \sum_{\alpha'=T',TL'}v_{\alpha'}{\cal R}^{\alpha'}\right] \, .
\label{sncspwia}
\ee
The hadronic response functions in PWIA are then given simply as the product of
the single-nucleon responses introduced above, ${\cal R}^\alpha$, and the (non-relativistic)
momentum distribution $N(p)$.

Although FSI and dynamical relativity (even in the plane-wave limit) in general destroy the factorization property, Eq.~(\ref{PWIA}) is still usually the basis for defining an
effective spectral function (also called reduced cross section) which is employed
to analyze and interpret experimental data.


\subsection{Polarized off-shell single-nucleon responses}


Within the plane-wave limit, the analysis of $(e,e'N)$ reactions involves
the half-off-shell $\gamma NN$ vertex. From parity and time reversal transformation 
properties together with Lorentz and gauge invariance, this may be shown to involve four form factors
that depend not only on $Q^2$ but also on the invariant mass $P^2$~\cite{NaPo90}. At present, a rigorous
approach to treating the off-shell dependences does not exist, and thus it is necessary to rely on
simple {\it ad hoc} prescriptions. This is the scheme initially developed by de 
Forest~\cite{For83} and later on generalized to more complex (polarized) 
situations~\cite{Cab93}. 

More specifically, the de Forest procedure involves the three following basic steps:
first, the spinors are treated as free (on-shell) ones; second, the current operator is chosen to be the on-shell one. Here two forms, called $CC1$ and $CC2$, have commonly been assumed
\ba
\Gamma^\mu_{CC1} &=& (F_1+F_2)\gamma^\mu-\frac{F_2}{2M_N}(\overline{P}+P_N)^\mu
\label{eq1s4} \\
\Gamma^\mu_{CC2} &=& F_1\gamma^\mu + \frac{iF_2}{2M_N}\sigma^{\mu\nu}Q_\nu \, ,
\label{eq2s4}
\ea
where $F_1$ and $F_2$ are the Pauli and Dirac form factors, respectively, that depend
only on $Q^2$, and the on-shell
variable $\overline{P}^\mu =(\overline{E},\np)$ with $\overline{E}=\sqrt{p^2+M_N^2}$ has
been introduced. Note that the $CC1$ operator is obtained from the $CC2$ one by using
the Gordon decomposition (only valid for free on-shell nucleons). 
In PWIA, however, the two current operators are in general different because of the
off-shell bound nucleon involved in the process. Finally,
the matrix elements of the operators $CC1$ and $CC2$ violate conservation of the
one-body current (impulse approximation).
Thus, six different possibilities to deal with 
the half-off-shell $\gamma NN$ vertex are commonly involved. They are connected with
the form of the current operator
selected, and how current conservation is imposed or not (choice of gauge). These six
off-shell prescriptions are denoted by: i) $NCC1$ and $NCC2$, where no current conservation is imposed
(Landau gauge), ii) $CC1^{(0)}$ and $CC2^{(0)}$, where current conservation is imposed by eliminating the longitudinal component (Coulomb gauge), and iii) $CC1^{(3)}$ and 
$CC2^{(3)}$, where current conservation is imposed by eliminating the time component (Weyl gauge).
One hopes that the differences found between the results obtained with these prescriptions may allow one to estimate
the uncertainty introduced by the off-shell effects in $(e,e'N)$ reactions, although it should always be kept in mind that these are merely prescriptions for off-shell behaviour, albeit popular ones.

Using trace techniques the
polarized recoil single-nucleon tensor introduced in Eq.~(\ref{PWIA}) is simply given by
\be
{\cal W}^{\mu\nu}=
\frac{1}{8M^{2}_{N}}Tr\left[\hat{\Gamma}^{\mu}(\pbarslash +M_{N})
\overline{\Gamma}^{\nu}(1+\gamma_{5}\snslash)(\pnslash +M_{N})\right] \, ,
\label{eq3s4}
\ee
where we have 
introduced the recoil nucleon spin projector \( (1+\gamma_{5}{\snslash}) /2 \) and use
the notation $\overline{\Gamma}^\mu =\gamma^0\Gamma^\mu\gamma^0$. From Eq.~(\ref{eq3s4})
it is clear that all antisymmetric contributions come from the $\gamma_5$ term, and thus
the single-nucleon tensor can be decomposed into its symmetric and antisymmetric parts
\be
{\cal W}^{\mu\nu}={\cal S}^{\mu\nu}+{\cal A}^{\mu\nu}(S_N) \, ,
\label{eq4s4}
\ee
with all of the dependence on the recoil nucleon polarization contained solely in the
antisymmetric tensor. When contracted with the leptonic tensor,
the symmetric term, ${\cal S}^{\mu\nu}$, gives rise to the electron-unpolarized 
single-nucleon responses, whereas the antisymmetric contribution, ${\cal A}^{\mu\nu}(S_N)$,
is to be contracted with the antisymmetric (polarized) term in the leptonic tensor. Thus, 
within PWIA the recoil nucleon polarization enters only in the electron-polarized
responses. Explicit expressions for the single-nucleon tensor ${\cal W}^{\mu\nu}$ for the
two forms of the current operator in Eqs.~(\ref{eq1s4},\ref{eq2s4}) are presented in Appendix A.

The various single-nucleon response functions that enter in $(\vec{e},e'\vec{N})$ reactions
are constructed directly as components of the single-nucleon tensor in Eq.~(\ref{eq4s4})
according to
\ba
{\cal R}^{L}&=&\left(\frac{q^2}{Q^2}\right)^2\left[{\cal S}^{00}-2\frac{\omega}{q}
{\cal S}^{03}+\frac{\omega^{2}}{q^{2}}{\cal S}^{33}\right]\label{eq5s4} \\
{\cal R}^{T}&=&{\cal S}^{22}+{\cal S}^{11} \label{eq6s4}\\
{\cal R}^{TL}&=&\frac{q^{2}}{|Q^{2}|}2\sqrt{2}\left[\cos \phi 
\left({\cal S}^{01}-\frac{\omega}{q}{\cal S}^{31}\right)
        -\sin\phi \left({\cal S}^{02}-\frac{\omega}{q}{\cal S}^{32}\right)
                \right] \label{eq7s4}\\
{\cal R}^{TT}&=&\left({\cal S}^{22}-{\cal S}^{11}\right)\cos 2\phi
        +2 {\cal S}^{12}\sin 2\phi \label{eq8s4}\\
{\cal R}^{T'}&=&-2{\cal A}^{12} \label{eq9s4}\\
{\cal R}^{TL'}&=&\frac{q^{2}}{Q^{2}}2\sqrt{2}\left[\sin\phi
\left({\cal A}^{01}-\frac{\omega}{q}{\cal A}^{31}\right)
+\cos\phi\left({\cal A}^{02}-\frac{\omega}{q}{\cal A}^{32}\right)\right] \, ,
\label{eq10s4}
\ea
where the labels $\{1,2,3\}$ refer to the three usual
directions that determine the nucleonic
plane (see~\cite{RaDo89,Cab93} for details). Note that
if gauge invariance is fulfilled, implying that ${\cal W}^{03}={\cal W}^{30}=
(\omega/q){\cal W}^{00}$ and ${\cal W}^{33}=(\omega/q)^2{\cal W}^{00}$, then 
the longitudinal and interference
transverse-longitudinal responses can be written uniquely in terms of the time
or longitudinal components alone. As mentioned above, in PWIA the current is not
conserved and thus the results depend on
how the responses are constructed from the tensor. 

Analytic expressions for the polarized single-nucleon responses corresponding to 
the $CC1$ and $CC2$ current operators and the Coulomb gauge, i.e., imposing current conservation
by substituting the longitudinal component in terms of the time component, are displayed in Appendix A . 
The explicit expressions for the unpolarized and target-polarized single-nucleon responses 
are given in~\cite{Cab93}.

Note that the normal contribution only enters in the interference $TL'$ response
for out-of-plane kinematics. In this work, all of the analysis 
is performed for co-planar kinematics, and thus only four responses are available within PWIA.
Finally, although explicit expressions for the responses are only displayed for the
two Coulomb gauge prescriptions (see Appendix A), in this work we also explore and present
results corresponding to the Landau and Weyl gauges with both choices of the current
operator.


\subsection{The `on-shell' prescription}


In~\cite{Ama96,Jesch} the exact on-shell operators for use between two-component spin
spinors were developed. As explained in the previous section, the usual procedure to
deal with off-shellness starts from current operators taken on-shell. However, when matrix 
elements are computed using bound initial wave functions, a basic
problem occurs that persists in the analysis of the off-shell $\gamma
NN$ vertex, namely, the breaking
of gauge invariance. Current conservation of the one-body current is
violated to a degree that depends on the
amount of off-shellness, characterized by the energy difference $(\omega-\overline{\omega})$. 
In this section we
consider an alternative approach, called the `on-shell' prescription, which avoids this kind of
ambiguity, but at the same time requires the physical missing energy to be replaced by an 
effective value. Hence, the `on-shell' prescription should be
simply considered as a different way to deal with the $\gamma NN$ vertex involved in
$(e,e'N)$ processes.

The basic idea involved in the `on-shell' approach is to force the bound 
nucleon to be on-shell.
This means that its energy and momentum are given by the free relation,
$E\equiv\overline{E}=\sqrt{p^2+M_N^2}$. From energy conservation in $(e,e'N)$ processes (see Section~2), 
the excitation energy $\cal{E}$ that results in this case is given by
\ba
{\cal{E}} &=& M_A-\sqrt{p^2+M_N^2}-\sqrt{p^2+M_B^{0^2}} \\
&=&-\left[E_S+\left(\sqrt{p^2+M_N^2}-M_N\right)+\left(\sqrt{p^2+M_B^{0^2}}-M_B^0
\right)\right]<0 \, ,
\label{eq14s4}
\ea
where $E_S\equiv M_N+M_B^0-M_A$ is the separation energy, i.e., 
the minimum energy needed to separate the nucleus $A$ 
into a nucleon and the residual nucleus $B$ in its ground state. Thus, the excitation energy and the bound nucleon momentum are no
longer independent, 
${\cal E}={\cal E}(p)$, and consequently one must be very careful to
be consistent when calculating all of the remaining kinematical
variables. Note that
the excitation energy as function of $p$ is proven to be negative for all $p$-values.
Its absolute value gets larger as $p$ increases, being equal to $-E_S$ for $p=0$.
Obviously, these results do not correspond to a physical
situation; the physical excitation energy, ${\cal{E}}\equiv E_B-E_B^0$, is defined in a way that requires it to be positive or zero.
It is important to remark that the unphysical result given by
Eq.~(\ref{eq14s4}) is consistent with the ambiguities introduced via the
different off-shell prescriptions where the excitation energy has been fixed to zero. In
such a case, only for $p=0$ and $E_S=0$ do the `on-shell' and off-shell prescriptions coincide
and no ambiguities appear. As the value of $p$ goes up, the excitation energy required to be
on-shell is negative (unphysical), and consequently, off-shellness
uncertainties appear, and 
the higher the $p$-value the more important they are.
Below we also present observables corresponding to the `on-shell' approach. 
How these results differ from
the off-shell ones may provide us with some estimate of the uncertainties inherent in the
treatment of the $\gamma NN$ vertex in $(e,e'N)$ reactions.



\subsection{Semi-relativistic reductions}


For a long time the standard procedure to treat $(e,e'N)$ reactions has
been based on a non-relativistic description of the hadronic current.
Bound nucleon wave functions have been described as solutions of the 
Schr\"odinger equation. Similarly, in DWIA a
non-relativistic treatment of FSI has been invoked in describing the ejected
nucleon wave function. Most nuclear models have been derived within such a 
non-relativistic framework, and hence, in order to be consistent with such 
descriptions of the states, one is also forced to perform some type of non-relativistic 
reduction of the relativistic electromagnetic current. 

The standard
procedure to derive non-relativistic expressions for the electromagnetic current
operators has been based on expansions in all of the independent dimensionless variables, i.e., energy and momentum transfer, $\lambda= \omega/2M_N$, $\kappa= q/2M_N$, and initial-state struck nucleon momentum,
$p/M_N$. Treating all of these as small is not justified in present
studies where $q$ and $\omega$ are  comparable to the nucleon mass. Hence, here we
treat the problem exactly for the transferred energy and momentum.
This analysis follows closely the
reductions already presented in some previous
papers~\cite{Ama96,Ama98,Jesch} --- again, as noted in the
introduction, these should really be called ``semi-relativistic''
approaches, since much of the relativistic content is not approximated.
In this
work, instead of making use of semi-relativistic expressions for the single-particle
current matrix elements, we directly expand the various (polarized and unpolarized) single-nucleon
responses starting from the fully-relativistic $CC1^{(0)}$ PWIA expressions.
Moreover, various semi-relativistic expansion alternatives are explored. 
Our main aim is to establish
how precise the semi-relativistic reductions are, and under which conditions and/or
for which observables they may or may not work. 

\begin{figure}
{\par\centering \resizebox*{1.0\textwidth}{0.6\textheight}
{\rotatebox{270}{\includegraphics{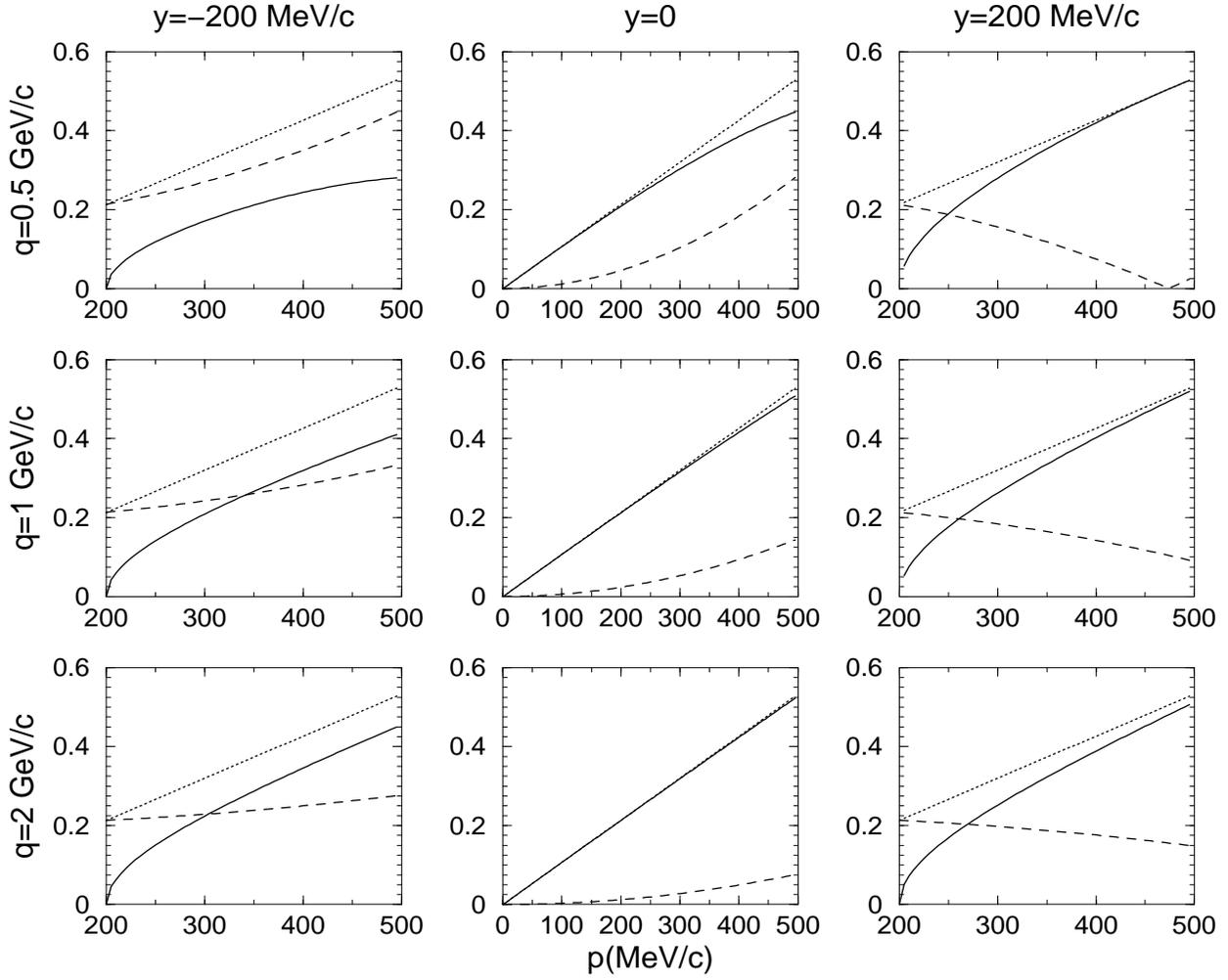}}}\par}
\caption{Behaviour of the kinematic variables $\chi$ (solid lines), $|\chi'|$ (dashed lines) and $\eta$ (dotted lines) as functions of the
missing momentum $p$ for various values of $q$ and $y$.\label{figure5}}
\end{figure}

Let us start by developing in detail the semi-relativistic expansion
procedures. We consider
three possible choices of the variable in which to make the semi-relativistic expansion:
$\chi\equiv (p/M_N)\sin\theta$, $\chi'\equiv (p/M_N)\cos\theta$ and
$\eta\equiv (p/M_N)$. The transferred energy and momentum are treated exactly.
In Fig.~\ref{figure5} we show the behaviour of $\chi$, $\chi'$ and $\eta$ as functions of 
the missing momentum $p$ for different kinematical situations defined
by $\{q,y,{\cal{E}}=0\}$ (see Section~2 for details on the kinematics). 
Throughout this work we present results over a wide range of values of $p$ --- the reader
should, of course, be aware that contributions other than those from the impulse
approximation may also be significant, especially for large $p$ (work is in progress to
provide relativistic treatments of such contributions).
From the results in Fig.~\ref{figure5} one may conclude that the best
variable in which to expand for the case $y\neq 0$ (away
from the QEP) is $\chi$. This is strictly valid up to a certain value of the
missing momentum $p$ where $\chi$ and $\chi'$ may cross each, after which $\chi'$ is the smaller, 
and hence more suitable as a choice of expansion variable. 
In the case of the quasielastic peak ($y=0$), 
the semi-relativistic reduction is expected to be more precise
if the expansion variable is chosen to be $\chi'$, instead of $\chi$ or $\eta$.
However, in order to estimate how valid a
semi-relativistic expansion might be, it is crucial to know where the $(e,e'N)$
cross section or hadronic response functions are mainly located. This
depends on the single-particle orbitals involved in the process.
In particular, for an $\ell =0$ shell the main location is centered around
$p=0$, whereas for $\ell = 1$ it is typically in the vicinity of $p\approx 100$
MeV/c. There one finds that $\chi$, $\chi'$ and
$\eta$ are all reasonably small and accordingly expansions are likely to be good.

In this work the semi-relativistic reduction of the responses is made by
expanding in powers of $\chi$.
There exist some general reasons to prefer $\chi$ over $\chi'$ or
$\eta$. By looking at the relativistic PWIA expressions for the polarized
single-nucleon responses in Eqs.~(\ref{rtprimel}-\ref{rtlprimen}), one realizes that $\chi$ enters as a
natural variable in those expressions. The same holds for the unpolarized
responses~\cite{Cab93}. 
Thus, the various polarized and unpolarized responses might be
classified into three basic categories: i) class ``0"
corresponding to responses which contain terms that begin with a finite leading-order contribution that does not depend on the
variable $\chi$, i.e. responses with finite terms proportional to $\chi^0$. This is
the case for ${\cal R}^L$, ${\cal R}^T$, ${\cal R}^{T'}_l$, ${\cal R}^{TL'}_s$
and  ${\cal R}^{TL'}_n$; ii) class ``1" responses whose leading-order contributions are linear in $\chi$, the case for ${\cal R}^{TL}$, ${\cal R}^{T'}_s$ and ${\cal R}^{TL'}_l$, and iii)
class ``2" responses with leading-order contributions proportional to $\chi^2$ as for ${\cal R}^{TT}$.
Below we shall examine how the effects of terms in the expansions that
are of higher order than the power that characterizes the class
influence the results, that is, whether or not a correlation exists
between the class and the relative importance of these higher-order
terms. In context it should also be mentioned that correlations with
class number can also be made when examining dynamical (versus
kinematical) relativistic effects~\cite{Cris1,Cab98a}.

Now let us proceed in detail to the semi-relativistic reduction of the
single-nucleon responses using $\chi$ as expansion variable.
We assume co-planar kinematics, i.e., $\phi=0$.
Following the discussion in Section~2, the $(e,e'N)$ process
is determined by four
`dynamical' variables. One is taken as the excitation energy
$\cal{E}$ fixed to zero, i.e., the residual nucleus is taken to be in its ground
state, and another is the variable $\chi$ used to perform the semi-relativistic
expansion. The two remaining `dynamical' variables
might be selected as $\{\kappa,\lambda\}$ or equivalently
$\{\kappa,y\}$. In this work, we explore the semi-relativistic expansion for
selected values of $\kappa$ and $y$. Note that all of the remaining
kinematical variables that may be introduced in the description of $(e,e'N)$
responses are given in terms of $\kappa$, $y$, $\cal{E}$ and  $\chi$.
Hence, within the semi-relativistic procedure developed in this work, they are
simply evaluated as a Taylor expansion in powers of $\chi$. 

After some algebra, keeping $\kappa$, $y$ and $\cal E$ constant (${\cal E}=0$ here), 
the semi-relativistic expressions for the
single-nucleon responses up to first order in powers of $\chi$ are given as follows
\ba
{\cal R}^{L}&=&\frac{\kappa^{2}}{\overline{\tau }_{y}}(F_{1}-
\overline{\tau}_{y}F_{2})^{2}+{\cal O}(\chi^{2}) \label{rlNR}\\
{\cal R}^{T}&=&2\overline{\tau}_{y}G_M^{2}+{\cal O}(\chi^{2}) \label{rtNR}\\
{\cal R}^{TL}&=&2\chi\sqrt{\frac{2\kappa ^{2}(1+\overline{\tau}_{y})}
{\overline{\tau}_{y}}}(F^{2}_{1}+\overline{\tau}_{y}F_{2}^{2})
\cos\phi +{\cal O}(\chi^{3}) \label{rtlNR} \\
{\cal R}^{TT}&=&-\chi^2(F_{1}^{2}+\overline{\tau}_{y}F^{2}_{2})
        \cos 2\phi +{\cal O}(\chi^{4}) \label{rttNR}\\
{\cal R}_{l}^{T'}&=&2\overline{\tau}_{y}G_M^{2}+
        {\cal O}(\chi^{2})\label{rtprimelNR} \\  
{\cal R}^{TL'}_{l}&=&2\sqrt{2}\chi\kappa a G_M        
        \left(eF_1+\sqrt{\kappa^2-\overline{\tau}_{y}}F_2\right)        
        \cos\phi+{\cal O}(\chi^{3}) \label{rtlprimelNR}\\ 
{\cal R}_{s}^{T'}&=&2\chi a
        G_M\left(\sqrt{\kappa^{2}-\overline{\tau}_{y}}F_{1}        
        -\overline{\tau}_{y}eF_{2}\right)+{\cal O}(\chi^{3}) 
                \label{rtprimesNR}\\ 
{\cal R}_{s}^{TL'}&=&2\sqrt{2}\kappa G_M\left(F_{1}-        
        \overline{\tau}_{y}F_{2}\right)\cos\phi+{\cal O}(\chi^{2}) \, , 
\label{rtlprimesNR}
\ea
where we use the dimensionless variables defined
in Appendix A and have introduced the term $\overline{\tau}_{y}$ that represents the value of
$\overline{\tau}$ evaluated at $p=|y|$. The factors $a$ and $e$ are defined as
\ba
a&\equiv&\frac{1}{\kappa+\sqrt{(1+\overline{\tau}_y)\left(\frac{\kappa^2}
{\overline{\tau}_y}-1\right)}} \label{eq:a} \\
e&\equiv&\sqrt{1+\left[\kappa+\sqrt{(1+\overline{\tau}_{y})
        \left(\frac{\kappa^2}{\overline{\tau}_y}-1\right)}\right]^2} \, .
\label{eq:e}
\ea
We do not show the semi-relativistic expression for the ${\cal R}^{TL'}_n$
response because it is absent in the case of co-planar kinematics.

Note that all of the single-nucleon responses have the following
generic form:
\begin{equation}
{\cal R}=\chi^n \alpha_0 [1+\alpha_2 \chi^2 + {\cal O}(\chi^4)] \, ,
\label{genexpan}
\end{equation}
where $n=0$, 1, 2 labels the class type, namely, the general expansions 
involve higher-order terms in $\chi^2$. The leading coefficient $\alpha_0$ is given in
Eqs.~(\ref{rlNR}-\ref{rtlprimesNR}) for all responses $L$, $T$, etc., and
depends on the details of the process (form factors, kinematic
factors, whether the nucleon is a proton or neutron). The relative
importance of the next-to-leading order contribution at small $\chi$
is characterized by the coefficient $\alpha_2$. To get some advance
insight into what will be discussed at length in the following
section, we provide a few typical numbers: in particular, we take
$q=1$ GeV/c, $y=0$ and focus on the region $p\leq 100$ MeV/c
(100 MeV/c is typical of valence proton knockout from $^{16}$O). We
truncate all expansions above at the terms involving $\alpha_2$ and
find the following results:
\begin{eqnarray}
\alpha_2(L,n=0) &=& 1.4     \nonumber \\
\alpha_2(T,n=0) &=& 1.2     \nonumber \\
\alpha_2(TL,n=1) &=& 0.3     \nonumber \\
\alpha_2(TT,n=2) &=& 0.1     \nonumber \\
\alpha_2(T'_l,n=0) &=& 1.1     \nonumber \\
\alpha_2(TL'_l,n=1) &=& -0.1     \nonumber \\
\alpha_2(T'_s,n=1) &=& 3.3     \nonumber \\
\alpha_2(TL'_s,n=0) &=& -1.4     
\label{valuesofalph}
\end{eqnarray}
For instance, for $p=100$ MeV/c one has $\chi^2=0.011$ and the relative
importance of the higher-order terms can be read off immediately. We
see that the class ``0'' cases have $\alpha_2$'s of order unity and
accordingly the next-to-leading order corrections are typically at the 1--2\%
level for $p=100$ MeV/c. The class ``1'' and class ``2'' cases are smaller by
typically a factor of a few (with the exception of the anomalous $T'_s$ case;
see below). Thus typically the coefficients $\alpha_2$ are
``natural'', meaning they are of order unity or somewhat
smaller. Clearly when $\chi$ is very small, such expansions should be
quite good, whereas when it rises to larger values (as it can) they will fail. The one anomaly is also easily explained,
since for it the leading-order coefficient $\alpha_0(T'_s)$ is not
``natural'', but from a cancellation of the expression in parentheses
in Eq.~(\ref{rtprimesNR}) is abnormally small, implying a larger than normal
role for the next-to-leading order term. In~\cite{Cris1} a similar
analysis is performed for dynamical relativistic effects to ascertain
the class pattern there as well.


\subsubsection*{\bf Comparison with other reduction schemes}


As we have mentioned, various alternatives can
be considered in making the semi-relativistic reductions of the responses. Here we briefly 
review some of them and compare with the results given in
Eqs.~(\ref{rlNR}-\ref{rtlprimesNR}). It is important to point out that these
alternative procedures are equally valid approaches, since there is really nothing
that favours one of them over another, apart from the comparison with the fully-relativistic
calculation.
\begin{itemize}
\item   In~\cite{Ama96,Ama98}, new `relativized' current
matrix elements were deduced by expanding only in powers of the bound nucleon
momentum $\eta\equiv p/M_N$, whereas in~\cite{Jesch} expansions were made in $\chi$, but
for the current matrix elements and
not for the single-nucleon response functions. Following the same procedure but at the level of
the single-nucleon responses instead of the on-shell current matrix elements, 
one can obtain on-shell expressions for the single-nucleon 
responses. In this case all responses are given simply in terms of $\chi$ and
$\overline{\tau}$ but no explicit dependence on $\kappa$ enters.
\item   Another alternative is to make the expansion in powers of 
$\chi$, as we did in the previous section, but using as independent variables
$\kappa$ and $\overline{\tau}$ instead of $\kappa$ and $y$. The results obtained are
formally identical to the ones given in Eqs.~(\ref{rlNR}-\ref{rtlprimesNR}), except 
for replacing  $\overline{\tau}_y$ by $\overline{\tau}$. This
semi-relativistic procedure 
also reduces to the previous one when the
value of $\kappa$ is approximated by $\kappa\approx
\sqrt{\overline{\tau}(1+\overline{\tau})}$.
\item Finally, in the case of the unpolarized responses we have also explored the results
obtained using the  `relativized' current matrix elements developed
in~\cite{Ama96,Jesch}.

\end{itemize}

Although not shown here for simplicity, we have studied the results obtained for the 
unpolarized and polarized single-nucleon --- actually single-proton ---
responses for a selected 
choice of kinematics where $q=500$ MeV/c, $y=0$ and ${{\cal E}=0}$
(for details on 
the kinematics, see the next section). We find that the results obtained by using the current 
matrix elements developed in previous studies~\cite{Ama96,Ama98,Jesch} are very similar 
to our semi-relativistic reductions except for 
${\cal R}^{TT}$. This last fact can be easily explained by realizing that ${\cal R}^{TT}$ is a
response of order $\chi^2$ (class ``2'') and the current matrix elements used in those studies
were deduced by expanding only up to first order in $\eta$. Obviously, terms of order
$\eta^0$ and $\eta^2$ should be considered in the matrix elements to construct the
${\cal R}^{TT}$ response.

We have also checked that the relative difference between the various
semi-relativistic reductions is in general smaller than the difference with respect to the
relativistic PWIA calculation. 
In the particular case of ${\cal R}^{TT}$ the various results almost coincide.
This comment also
applies to the responses ${\cal R}^{TL}$ and ${\cal R}^{TL'}_l$ where the
differences between the various relativistic and semi-relativistic calculations are tiny. 
On the contrary, the effect introduced by the
choice of the semi-relativistic expansion is more important for 
${\cal R}^L$, ${\cal R}^T$, ${\cal R}^{T'}_l$ and ${\cal R}^{TL'}_s$. Nevertheless, we 
conclude that the various alternative semi-relativistic
expansions explored produce results which are rather
similar, the discrepancy being at most of the order of $\sim$6--7\%
even for $p=500$ MeV/c. Hence in what follows we present a systematic
analysis of the kinematical relativistic effects by comparing the
semi-relativistic results obtained by making an expansion in powers of $\chi$
fixing $\kappa$, $y$ and $\cal{E}$ (see
Eqs.~(\ref{rlNR}-\ref{rtlprimesNR})) with the relativistic `off-shell' PWIA
results, as well as with the `on-shell' prescription.


\section{Results}


In this section we present the results obtained for the single-nucleon response functions
and transferred polarization asymmetries.
Our main aim is to establish the importance of the `off-shell' uncertainties and
kinematical relativistic effects. Accordingly we compare results corresponding to various
`off-shell' prescriptions with the calculations evaluated within the `on-shell' approach,
as well as with the semi-relativistic reduction developed in the previous section. A systematic
study is presented by analyzing different kinematical situations.

Let us recall that 
four `dynamical' 
variables, whose preferred choice depends on the specifics of the $(e,e'N)$ reaction, 
determine the response functions. In this work we assume co-planar kinematics,
i.e., $\phi=0$, and the excitation energy is taken ${\cal{E}}=0$. Single-nucleon
responses and polarization asymmetries are shown as functions of the missing momentum
$p$. We consider only the proton case, $(e,e'p)$. 

\begin{figure}
{\par\centering \resizebox*{0.65\textwidth}{0.65\textheight}
{\includegraphics{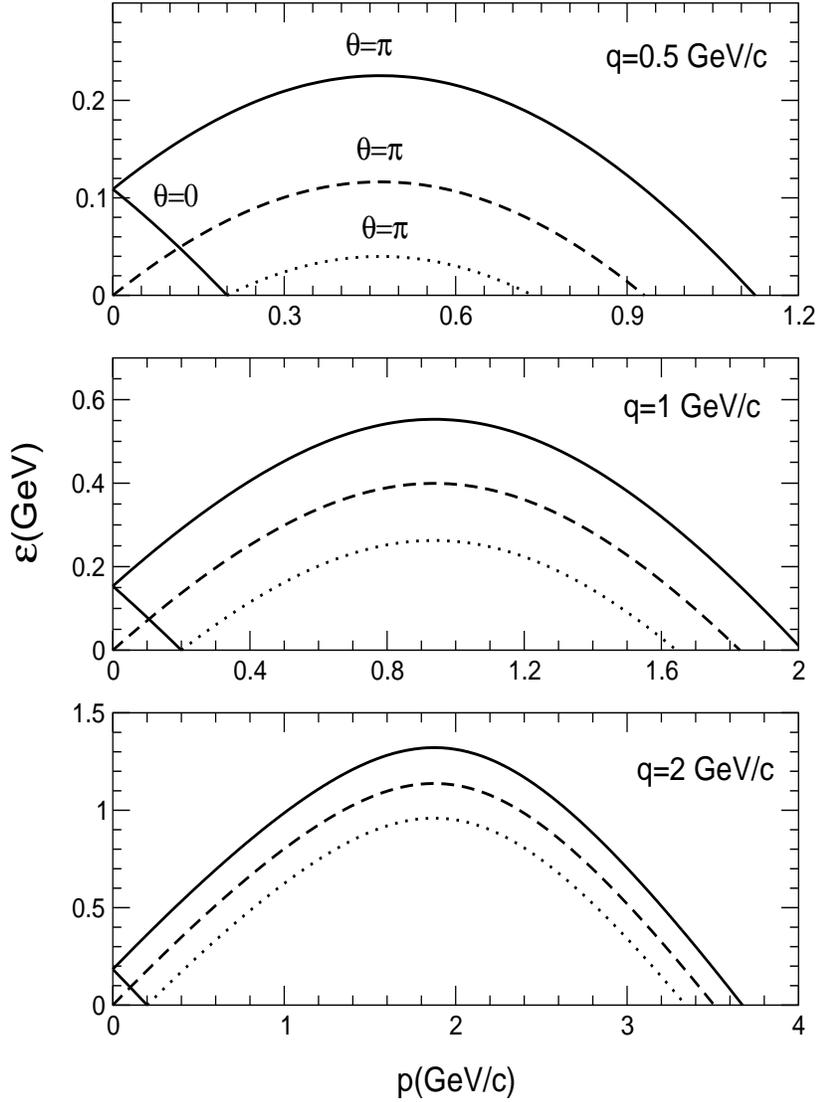}} \par}
\caption{\label{figure7}Excitation energy as function of $p$ for
$^{16}$O. Curves are plotted for $q=$ 0.5, 1 and 2 GeV/c and
$y=200$ MeV/c (solid line), $y=0$ (dashed line) and $y=-200$ MeV/c (dotted
line).} 
\end{figure}

Two different choices of kinematics have been selected:
\begin{itemize}
\item   Kinematics specified by the `dynamical'
        variables $\{q,y,{\cal{E}},p\}$, i.e., the observables are evaluated for selected
        values of $q$ and $y$. In Fig.~\ref{figure7} we show the allowed 
        $({\cal{E}}, p)$ regions for various values of $q$ and $y$. 
        Note that the allowed $p$-region is different for each $\{q,y\}$ choice. 
        The figure is intended only to show the complete range of kinematics allowed and thus
        to make clear the entire behaviour of the variables involved. Again, as noted above,
        the reader is cautioned about expecting the impulse approximation to provide a complete
        description for the very large values of $p$ covered in the figure.
        We recall that for the off-shell prescriptions these kinematics are equivalent 
        to the more popular $(q-\omega)$-constant kinematics, the corresponding value 
        of $\omega$ for each $y$ being calculated by means of Eq.~(\ref{eq:w(q,y)}).  
\item   Parallel kinematics, i.e., the polar angle $\theta_N$ is fixed to zero (see Section 2).
        Two alternatives
        have been considered. First, the kinetic energy of the outgoing nucleon is fixed; this
        means that the process is described by the dynamical variables 
        $\{p_N,\theta_N,{\cal{E}},p\}$.
        Here note that varying the missing momentum $p$ means changing the momentum transfer $q$.
        Second, the momentum transfer is fixed so that $\{q,\theta_N,{\cal{E}},p\}$
        constitute the four dynamical variables. In this case, one must
        change the outgoing nucleon momentum $p_N$ in order to vary the
        momentum $p$. It is important to point out that in both of these
        cases the value of $y$ coincides with $p$, i.e., $p=|y|$ and hence
        increasing $p$ means moving far away from the quasielastic peak.
        Within parallel kinematics, the angle 
        $\theta$ between $\nq$ and $\np$ might be $0$ (sometimes 
        referred as positive $p$-region or strictly parallel kinematics) 
        or $180^0$ (negative-$p$ or antiparallel kinematics). In the   
        first case, $\theta=0$, one has $p_N > q$ which corresponds to       
        $y>0$. In this situation, note that as $p$ increases the transfer
        momentum $q$ diminishes (for $p_N$-fixed) or 
        analogously, the outgoing nucleon momentum $p_N$ goes up         
        (for $q$-fixed). For $\np$ antiparallel to $\nq$, i.e., 
        $\theta=180^0$, one has $p_N<q$ implying $y$-negative. Here,
        increasing $p$ means also increasing $q$ ($p_N$-fixed) or 
        diminishing $p_N$ ($q$-fixed).

\end{itemize}

\subsection{Analysis of single-nucleon responses for $\{q,y\}$-fixed kinematics}


We start by showing the responses corresponding to $\{q,y\}$-fixed kinematics. 
In Figs.~\ref{figure8}-\ref{figure11} we present
the four unpolarized responses, and likewise in Figs.~\ref{figure12}-\ref{figure15} 
the four recoil nucleon
polarized responses that enter in co-planar kinematics. For each
response we present the results for three values of the momentum transfer:
$q=0.5$ GeV/c (top panels), $q=1$ GeV/c (middle panels) and $q=2$ GeV/c (bottom panels). 
Three different values of $y$ have been analyzed: 
$y=-200$ MeV/c ($\omega<\omega_{QEP}$) (left panels), $y=0$ (at the quasielastic peak)
(middle panels) and $y=+200$ MeV/c ($\omega>\omega_{QEP}$) (right panels). 
This provides a sampling of different situations when trying to understand how the
`off-shell' and relativistic effects enter in the various responses and how they vary
with momentum transfer and location with respect to the QEP. For the gauge-dependent 
($L$ and $TL$) responses we
present in each individual graph five curves that correspond to the three different
`off-shell' prescriptions based on the use of the $CC2$ current, 
the `on-shell' approach (dotted line) and the 
semi-relativistic reduction (long-dashed line) developed in the previous section. 
Within the off-shell curves we distinguish the three gauges: Coulomb
($CC2^{(0)}$) (solid lines), Landau ($NCC2$) (dot-dashed)
and Weyl ($CC2^{(3)}$) (short-dashed). Although not shown in these graphs (for clarity), 
we have also studied the behaviour of
the prescriptions based on the $CC1$ current, and we
comment on those results in the discussion that follows. On the contrary, in the case of 
the purely transverse responses (no gauge ambiguities)
we show the results obtained for the two current operators: $CC1$ (thin solid lines) and
$CC2$ (thick solid lines).

\subsubsection*{Unpolarized responses}

\begin{figure}
{\par\centering \resizebox*{0.8\textwidth}{0.8\textheight}
    {\includegraphics{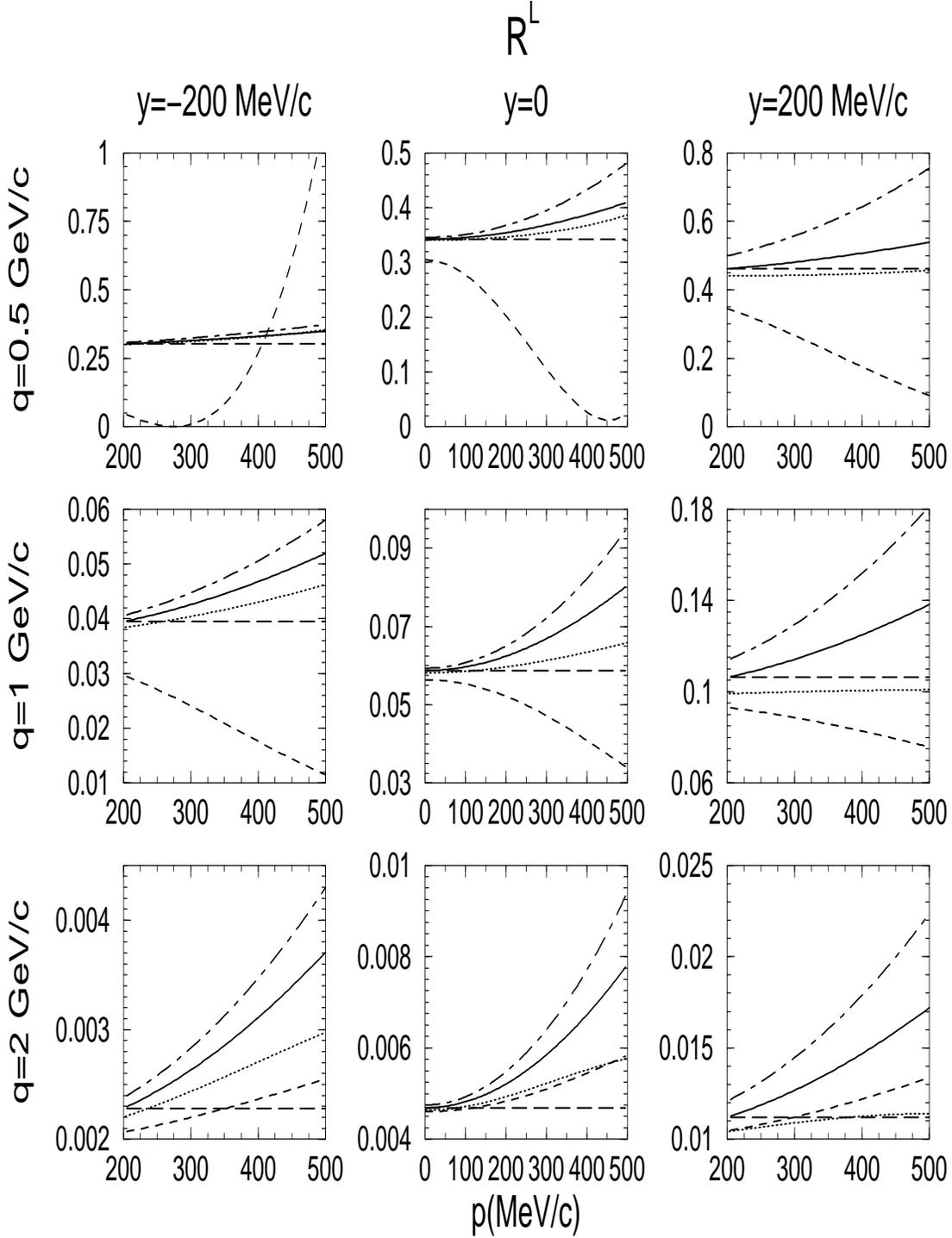}} \par}
\caption{\label{figure8}Unpolarized longitudinal single-nucleon response
${\cal R}^L$ for $\{q,y\}$-fixed kinematics. The excitation energy $\cal{E}$ is taken to be
zero. For the $CC2$ off-shell prescriptions the three gauges considered are: Coulomb (solid lines),
Landau (dot-dashed lines) and Weyl (short-dashed lines). The `on-shell' approach is
represented by dotted lines and the semi-relativistic reduction by long-dashed lines.}
\end{figure}

\begin{figure}
{\par\centering \resizebox*{0.8\textwidth}{0.8\textheight}
    {\includegraphics{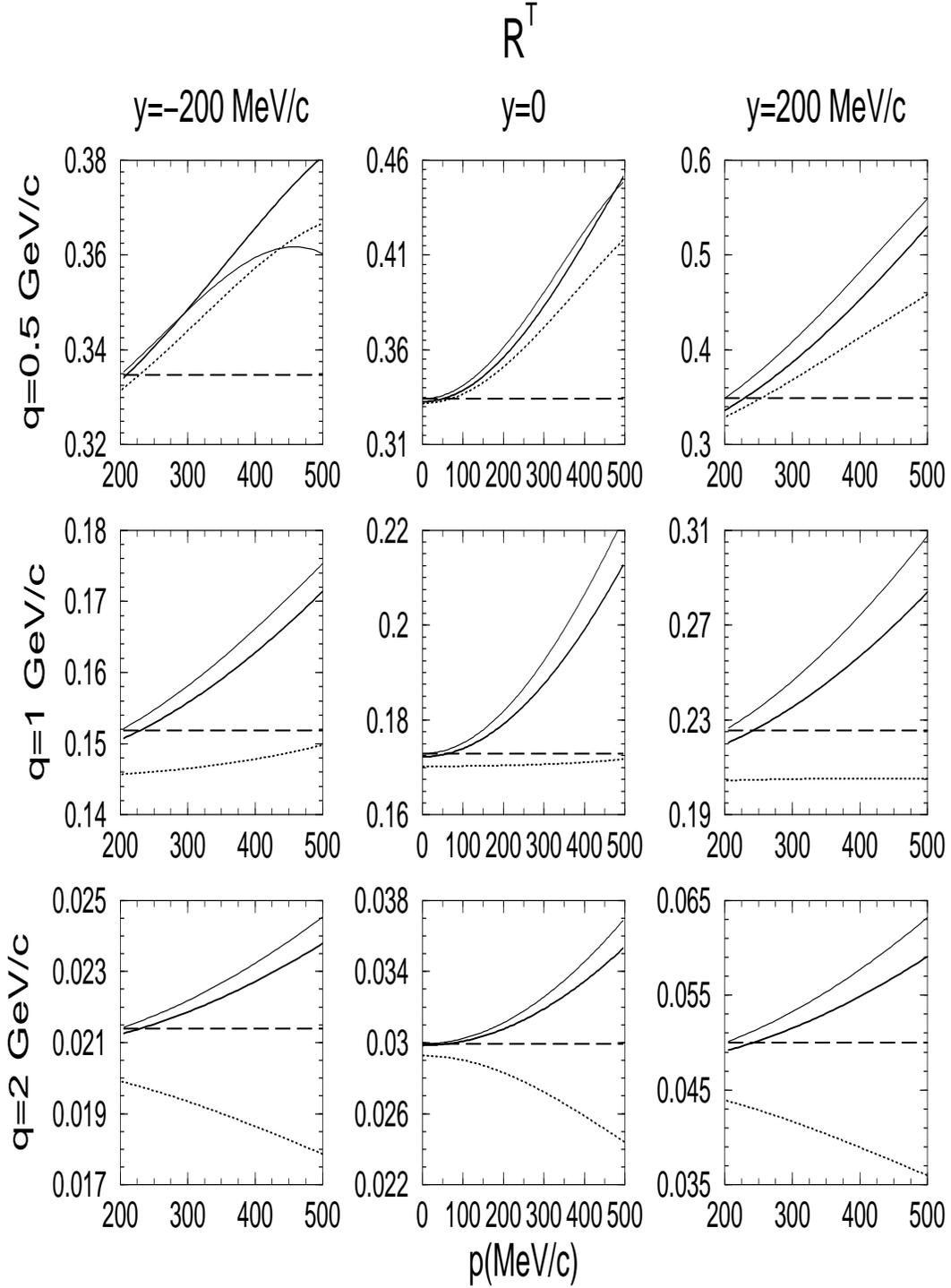}} \par}
\caption{\label{figure9}Same as Fig.~\ref{figure8}, but for the ${\cal R}^T$ response. We represent also the results for the $CC1$ current with thin solid lines.}
\end{figure}

\begin{figure}
{\par\centering \resizebox*{0.8\textwidth}{0.8\textheight}
    {\includegraphics{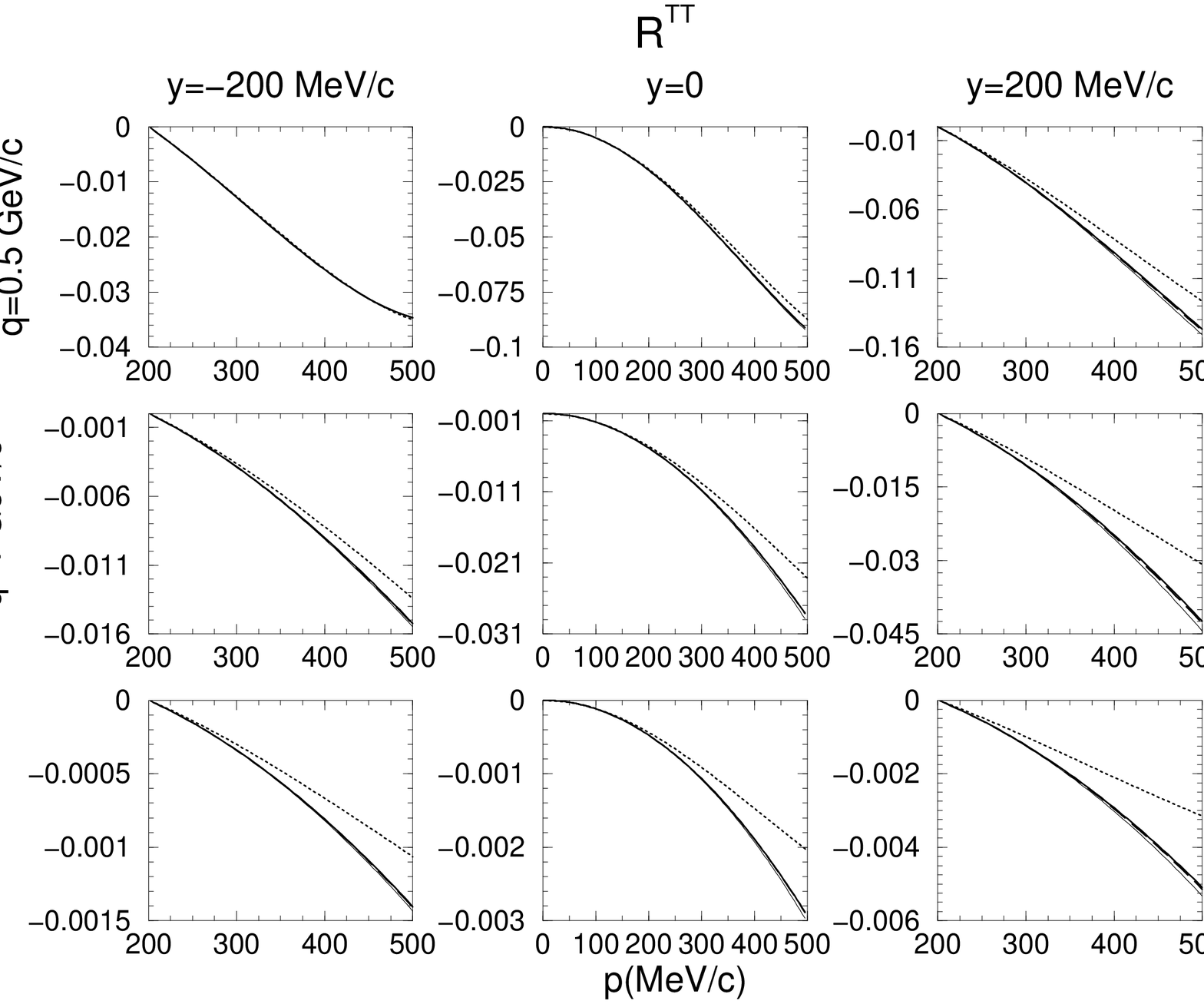}} \par}
\caption{\label{figure10}Same as Fig.~\ref{figure9}, but for ${\cal R}^{TT}$.}
\end{figure}

\begin{figure}
{\par\centering \resizebox*{0.8\textwidth}{0.8\textheight}
    {\includegraphics{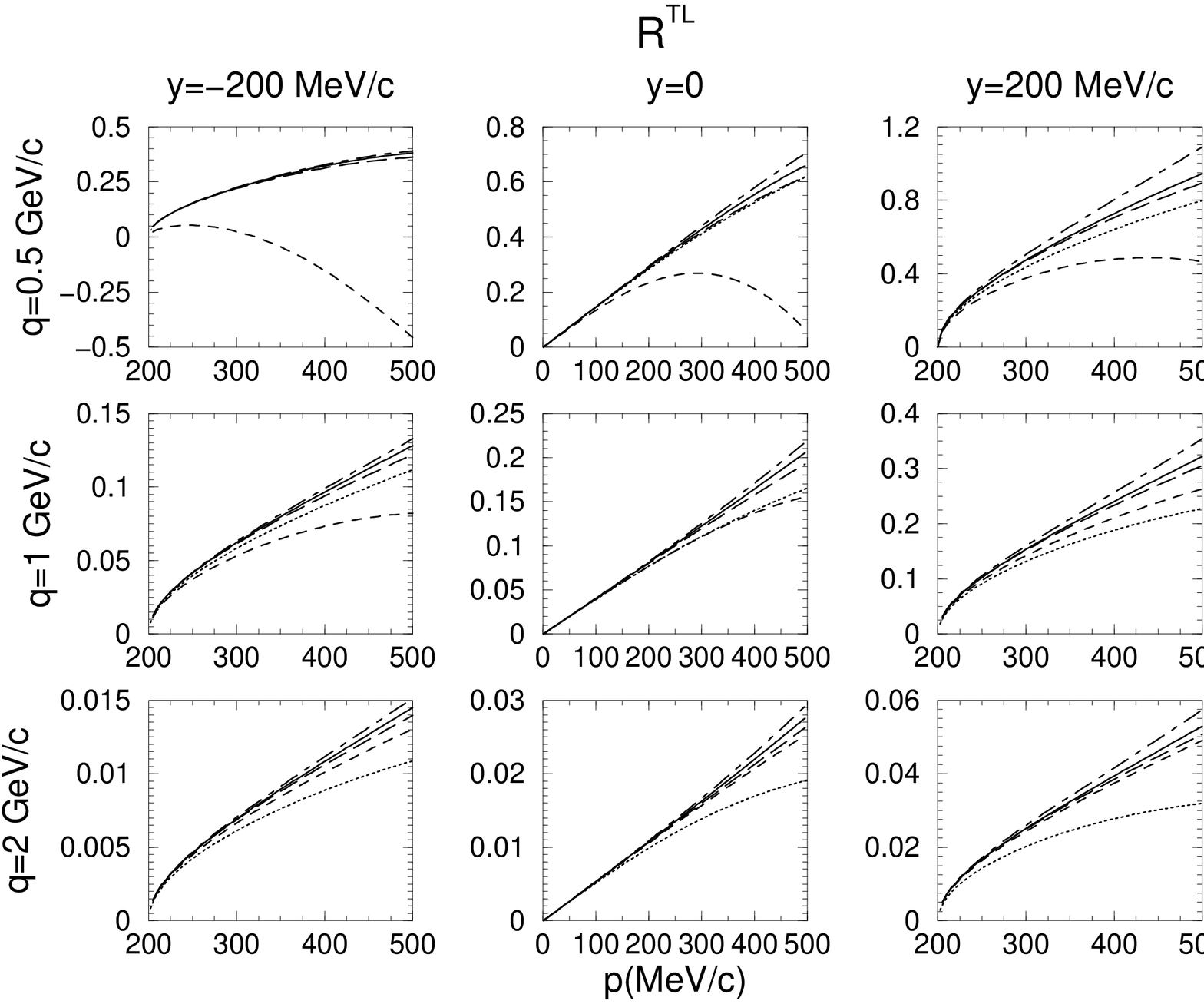}} \par}
\caption{\label{figure11}Same as Fig.~\ref{figure8}, but for ${\cal R}^{TL}$.}
\end{figure}

Let us start with the analysis of the off-shell effects in
the unpolarized responses (Figs.~\ref{figure8}-\ref{figure11}). 
First, we discuss the ambiguities introduced by the current operator choice. 
Note that although these effects are only shown in the graphs for the purely transverse 
responses ${\cal R}^T$ and ${\cal R}^{TT}$ (which only depend on the
choice of the current operator), they have been also analysed for ${\cal R}^L$ and 
${\cal R}^{TL}$ which depend additionally on the choice of gauge. Focusing on the Coulomb 
gauge, the effects introduced by the
current operator choice are proven to be tiny for all of the responses and all of the $\{q,y\}$-values
selected, even for high missing momenta. In particular, the response 
${\cal R}^L$ does not depend on the current choice~\cite{Cab93}, while the
Gordon uncertainties are at most of the order of $\sim$5\% for ${\cal R}^{TL}$ and
${\cal R}^{TT}$, and less than $\sim$8--9\% for ${\cal R}^T$. 
A general dependence of the Gordon uncertainties with
the energy transfer and momentum cannot clearly be extracted for the four responses. A similar
discussion could be also applied to the Landau gauge. The
uncertainty in this case is at most of the order of $\sim$5--6\%, being a maximum for
$y=200$ MeV/c.
In contrast, in the case of the Weyl gauge, the uncertainty associated with the current 
operator choice is much bigger, although it diminishes as $q$ increases (assuming $y$-fixed).


In what follows we analyze the uncertainty due to the choice of gauge.
As is well-known, this issue only arises for the ${\cal R}^L$ and ${\cal R}^{TL}$
responses. The results in Figs.~\ref{figure8} and \ref{figure11} 
show that the largest differences are
introduced by the Weyl gauge. This agrees with the
general findings in~\cite{Cris1,Cab98a,Cab93}. On the other hand,
the gauge ambiguities vary significantly with the specifics of the kinematics, i.e., the $q$- and $y$-values selected. In particular, the following general
trends are observed:
\begin{itemize}
\item   For $q$-fixed, the Weyl results tend to approach the results for the other two
        gauges as one goes from negative to positive $y$-values. This
        behaviour shows up particularly for small $q$. For $q=2$
        GeV/c the gauge deviations do not show any appreciable change when
        varying $y$. Moreover, the gauge uncertainties diminish significantly as $q$ increases. 
        Notably, this decrease is proven to be much faster for negative-$y$.
\item   The relative deviations between
        the Coulomb and Landau gauges
        increase as one moves to higher $\omega$. The
        magnitude of this increase is stronger for small $q$-values.
\end{itemize}

To finish the discussion of the off-shell prescriptions and in order
to understand the origin and significance of the relative effects discussed
above, it is crucial to point out that the $p$-range for which the responses
have been evaluated, $\leq 500$ MeV/c, may represent for each $\{q,y\}$
choice a very different fraction of the total kinematically allowed
$p$-region. In fact, as illustrated in Fig.~\ref{figure7}, the allowed $p$-region
is much wider as $q$ increases. This means that $|y|\leq p\leq 500$ MeV/c
may represent $\sim$55\% of the total $p$-region for $q=500$
MeV/c (for $y=200$ MeV/c, $\sim$30\%), 
while it only represents about $\sim$9--10\% for $q=2$
GeV/c. This might explain why the gauge differences discussed above are
smaller for higher $q$.

Next we turn briefly to the `on-shell' approach. We include it here
for several reasons: (1) it has been used in other 
studies~\cite{Ama96,Ama98,Jesch} and
it is important to see how it agrees with or differs from the popular
off-shell prescriptions; (2) it has the merit of providing a clear
connection between current matrix elements (with corresponding current
operators) and single-nucleon cross sections; and (3) it has the nice
feature that current conservation is built in from the start and does
not have to be imposed in the somewhat artificial way it does for the
de Forest $CC$ prescriptions.

The results corresponding to the `on-shell' prescription are represented by dotted
lines. As is seen in the figures, its deviation with respect to the off-shell prescriptions
increases as the momentum and energy transfers
grow (from negative to positive $y$-values in the latter case). This behaviour is strictly valid for the 
${\cal R}^T$ and ${\cal R}^{TT}$ responses where gauge ambiguities do not enter, and
also for the prescriptions based on the Coulomb and Landau gauges in the case of
${\cal R}^L$ and ${\cal R}^{TL}$. As noted above, the Weyl results at low $q$ differ from all other cases considered, including the `on-shell' prescription.

In order to clarify the comparison between the
various off-shell prescriptions and the `on-shell' approach, let us state explicitly the approach taken. First, the
four dynamical variables that completely determine the single-nucleon responses in general are given by $\{q,y,{\cal{E}},p\}$. For
the off-shell approaches the excitation energy is fixed (${\cal{E}}=0$ in this work), and once the dynamical
variables $\{q,y,{\cal{E}}\}$ are given, the energy transfer $\omega$ can be evaluated. It is also usual to introduce an
`on-shell' transferred energy defined as $\overline{\omega}\equiv
E_N-\overline{E}$, where $\overline{E}$ is the on-shell nucleon energy, which depends only on $p$. For the `on-shell' approach we fix $\{q,y,p\}$, determining $\cal E$ via Eq.~(\ref{eq14s4}) (being forced
to be an explicit function of $p$), then calculating the effective energy transfer $\omega_{on-shell}$ by using Eq.~(\ref{eq:w(q,y)}). It can be proven that
the two on-shell transfer energies $\overline{\omega}$ and
$\omega_{on-shell}$ do not coincide exactly; however their difference can be shown to be almost
negligible for all $p$ (at most of the order of $0.08$ MeV in the case of the 
residual nucleus $^{15}$N).

The reasons for the differences amongst the results based on the `on-shell' and
on the off-shell prescriptions can be traced back to the different energy transfers employed in each case. They can affect the results in different ways:
\begin{itemize}
\item For the off-shell prescriptions, the nucleon form factors $F_1$ and $F_2$ are 
functions of the real four-momentum transfer, i.e. $F_{1,2}=F_{1,2}(Q^2)$ with $Q^2=\omega^2-q^2$. 
However, for the `on-shell' prescription the dependence of these form factors is taken to be 
$Q^2_{on-shell}=\omega^2_{on-shell}-q^2$. The dipole $\tau$ dependence of the electric and magnetic
form factors considered in this work, explains why the discrepancy between
$F_{1,2}(Q^2)$ and $F_{1,2}(Q^2_{on-shell})$ increases for higher values of
$q$ and $y$. An alternative `on-shell' prescription (not considered here) is to employ form factors
$F_{1,2}(Q^2)$ in place of  $F_{1,2}(Q^2_{on-shell})$  with all else left unchanged. 
\item In the case of the Landau and Weyl gauges,
because of the way of
imposing current conservation (Weyl gauge) or not enforcing it (Landau),
an additional dependence on the off-shell energy transfer $\omega$ enters in
the single-nucleon responses. These are evaluated as 
${\cal R}^L=(q/\omega)^2{\cal S}^{33}$ and 
${\cal R}^{TL}=2\sqrt{2}(q/\omega){\cal S}^{31}$ (coplanar kinematics)
for the Weyl gauge, and as given by Eqs.~(\ref{eq5s4},\ref{eq7s4}) for the Landau gauge. 
For the `on-shell' prescription, current conservation is fulfilled, and this 
additional dependence on $\omega$ does not enter.
Obviously, in the case of the $CC2$ choice of the current such
$\omega$ dependence is also contained within the current operator itself. 
\end{itemize}

To finish the discussion of the unpolarized single-nucleon responses
let us consider the results corresponding to the semi-relativistic reduction
(long-dashed lines). We recall that any comments made for the single-nucleon responses can 
be extended to the hadronic responses (when we neglect contributions from negative-energy projections), 
since the relativistic and non-relativistic momentum distributions are quite similar 
over the whole momentum range considered~\cite{Cab98a}. As shown, it agrees nicely with the relativistic
calculations for the two interference responses,
particularly for ${\cal R}^{TT}$. The discrepancy
is at most of the order of $\sim$6--8\% for the high-$p$ region, where
the semi-relativistic reduction 
is less successful. The situation is clearly
different for ${\cal R}^L$ and ${\cal R}^T$ where the divergence between
relativistic and semi-relativistic results might be even of the order
of $\sim$40\% 
for some specific kinematics for the higher $p$-values. Moreover, the semi-relativistic responses
are almost constant over the entire range of $p$-values. This behaviour can easily be traced back by looking at the semi-relativistic expressions in Eqs.~(\ref{rlNR},\ref{rtNR}). The
variation with $p$ contained in the relativistic responses is due to the order-$\chi^2$
terms entering in the relativistic expressions~\cite{Cab93}. Finally, it
is hard to deduce any clear systematics concerning the quality of the semi-relativistic
reduction, and how it improves or gets worse with the momentum and
energy transfers. What one can deduce is that, while a reasonable
approximation for small missing momenta (say on the order of 100 MeV/c
or smaller), such a semi-relativistic description does not seem to be 
adequate to describe the ${\cal R}^L$ and ${\cal R}^T$ responses at
higher values of $p$. Even for intermediate $p$-values ($200\leq p\leq
300$ MeV/c) the semi-relativistic reductions fail to some degree, 
as the deviations can be already of the order of $\sim$15--20\%.

\subsubsection*{Polarized responses}

\begin{figure}
{\par\centering \resizebox*{0.8\textwidth}{0.8\textheight}
{\includegraphics{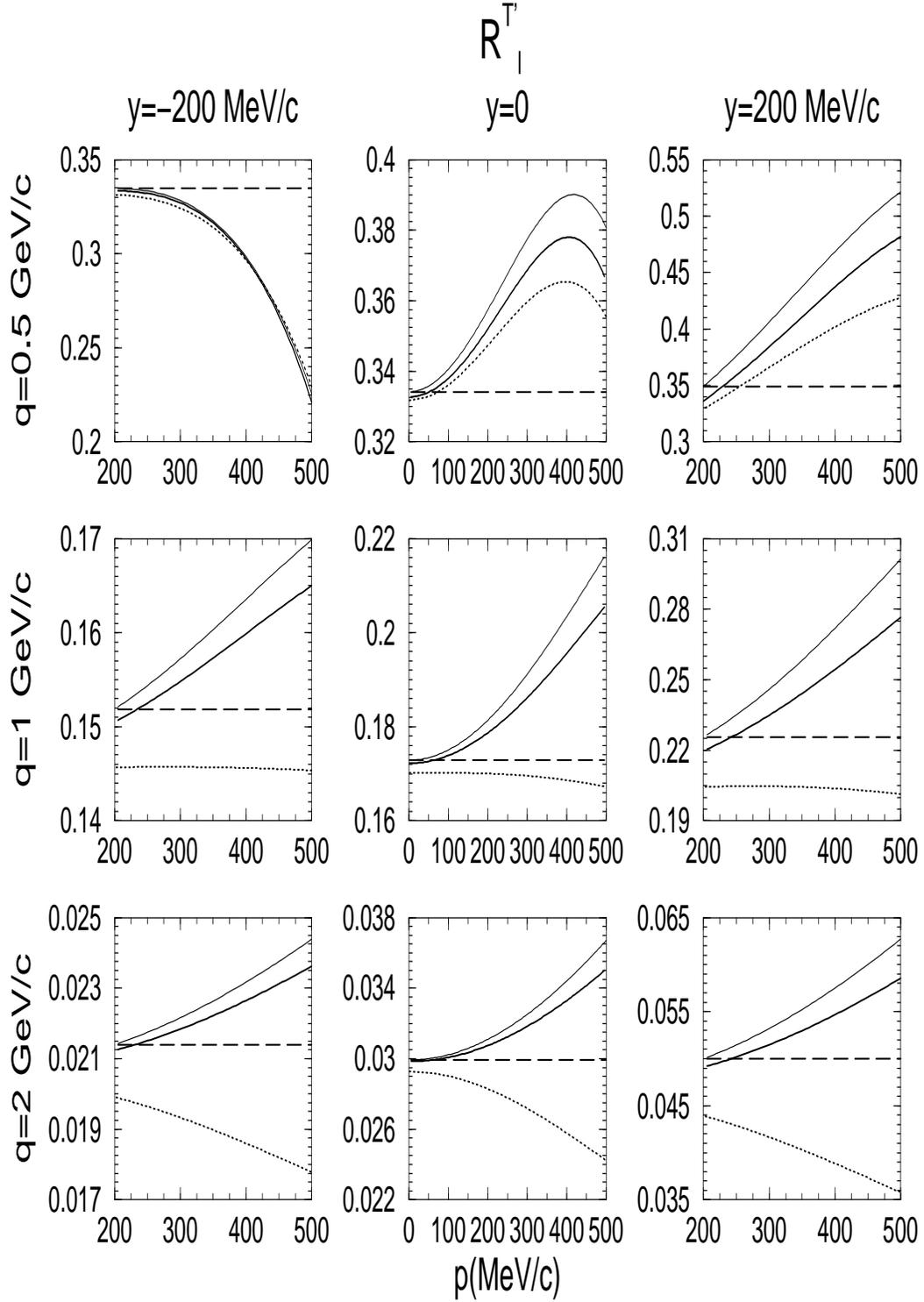}} \par}
\caption{\label{figure12} Same as Fig.~\ref{figure9}, but for the polarized response ${\cal R}^{T'}_l$.}
\end{figure}

\begin{figure}
{\par\centering \resizebox*{0.8\textwidth}{0.8\textheight}
{\includegraphics{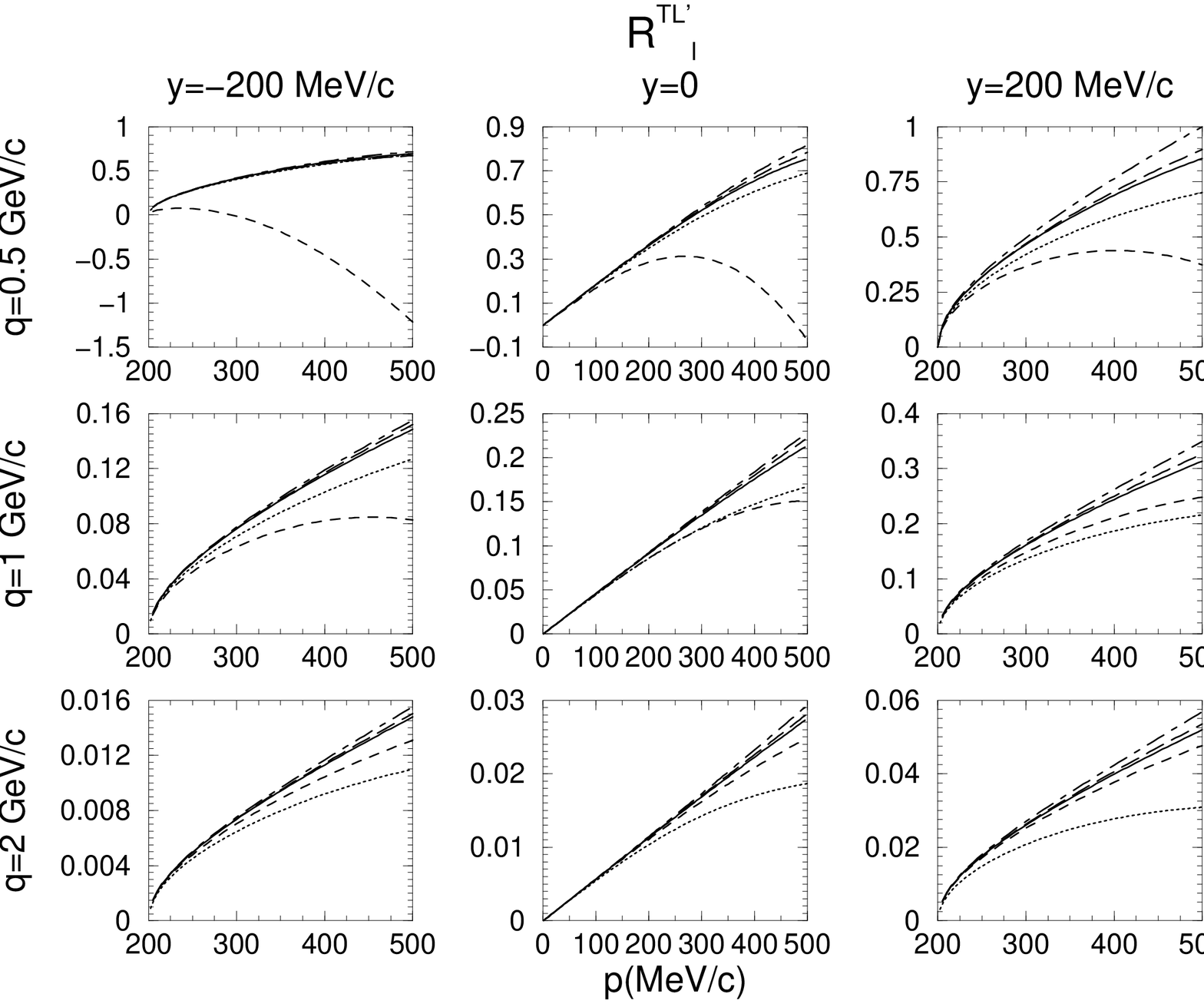}} \par}
\caption{\label{figure13}Same as Fig.~\ref{figure8}, but for the response ${\cal R}^{TL'}_l$.}
\end{figure}

\begin{figure}
{\par\centering \resizebox*{0.8\textwidth}{0.8\textheight}
{\includegraphics{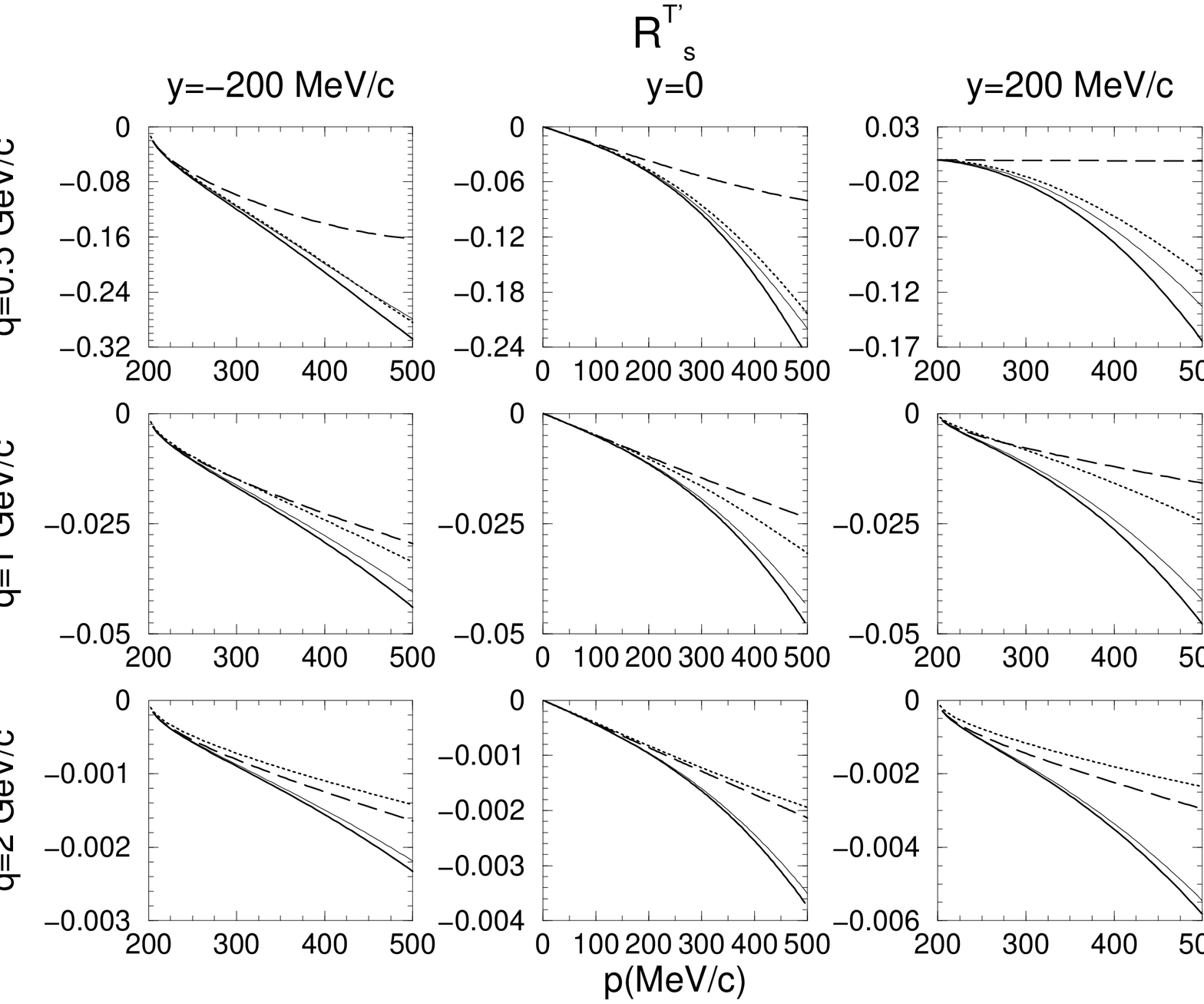}} \par}
\caption{\label{figure14}Same as Fig.~\ref{figure9}, but for the response ${\cal R}^{T'}_s$.}
\end{figure}

\begin{figure}
{\par\centering \resizebox*{0.8\textwidth}{0.8\textheight}
{\includegraphics{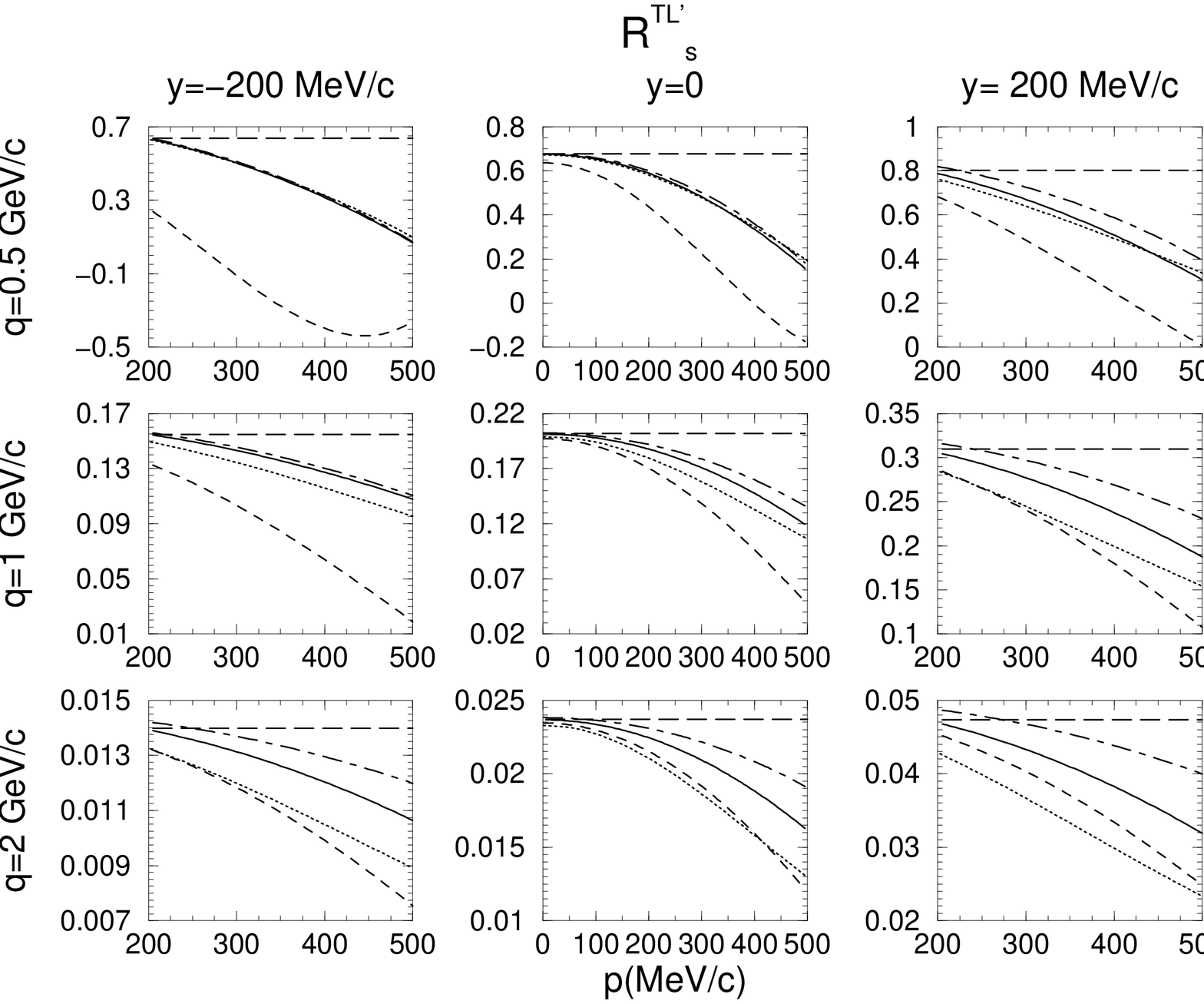}} \par}
\caption{\label{figure15}Same as Fig.~\ref{figure8}, but for the response ${\cal R}^{TL'}_s$.}
\end{figure}

The four recoil nucleon polarized responses that enter in
the analysis of $(\vec{e},e'\vec{p})$ reactions within PWIA and co-planar kinematics
are shown in Figs.~\ref{figure12}-\ref{figure15}.
The labels are the same as for the unpolarized responses.
Some of the discussion already presented above for the unpolarized responses can be also applied
to these polarized ones. Let us start by analyzing the uncertainty due to the choice
of the current operator. As shown in the figures, in the case of the two purely transverse responses,
${\cal R}^{T'}_l$ and ${\cal R}^{T'}_s$, which do not depend on the gauge, the
uncertainty is rather small
for all of the $\{q,y\}$-values selected. In the case of the interference $TL$
responses (not shown), ${\cal R}^{TL'}_l$ and ${\cal R}^{TL'}_s$, we find that
the Gordon ambiguity is also typically, but with some exceptions, relatively small for the Coulomb
and Landau gauges. The same result holds for the
Weyl gauge in ${\cal R}^{TL'}_l$, while the discrepancy introduced in
${\cal R}^{TL'}_s$ is significantly enhanced, in particular for large $p$-values.

The discussion of the gauge uncertainties follows similar trends to the ones presented
for the unpolarized responses. Summarizing the basic results, one may conclude
that the spread in the curves evaluated with the Coulomb and Landau gauges tends to be wider
as $y$ increases. On the contrary, the Weyl gauge results tend to come closer, as 
well as they do as $q$ increases.

Regarding the `on-shell' prescription (dotted line), the discussion follows the same
trends already presented at length in the case of the unpolarized responses, and therefore
we will not repeat them here. The reader is referred to the above discussion for details.
We should only point out that the `on-shell' approach gives very similar results to the
off-shell ones in the case
of small/medium $q$ and negative $y$ (see the results for $q=500$ MeV/c and $y=-200$ MeV/c). An exception
is the Weyl gauge whose results and general behaviour, as already discussed, 
can be very different from Coulomb and Landau gauge results.

Finally, with regard to the behaviour of the semi-relativistic reductions for the
polarized responses (long-dashed lines), the following conclusions can be drawn:
\begin{itemize}
\item   The semi-relativistic reduction works nicely for ${\cal R}^{TL'}_l$. In fact, this
  polarized response is the one that shows the smallest kinematical relativistic effects.
  In particular, the relative difference between the semi-relativistic results 
  and the $CC1^{(0)}$ prescription
  is less than $\sim$7\% even at high missing momenta. This behaviour echoes the one already
  observed for the unpolarized ${\cal R}^{TL}$ and ${\cal R}^{TT}$
  responses. It is also interesting to point out the similarity between 
  ${\cal R}^{TL}$ and ${\cal R}^{TL'}_l$, particularly for large $q$, independent of the approach considered.
\item   The kinematical relativistic effects in the response ${\cal
  R}^{T'}_l$ are in general quite high for large missing momenta (for
  some particular kinematics they can be of the order of
  $\sim$20--30\%). Only in the low-$p$ region ($p < 200$ MeV/c) is a
  semi-relativistic reduction for this response acceptable, as the
  differences are less than $\sim$6\% (for $|y|=200$ MeV/c this is
  true up to $250$ MeV/c). This result is similar to the one seen for ${\cal R}^L$
  and ${\cal R}^T$. Note that the semi-relativistic expressions for ${\cal R}^T$ and
  ${\cal R}^{T'}_l$ are identical (see Eqs.~(\ref{rtNR},\ref{rtprimelNR})). Moreover, the 
  off- and `on-shell' relativistic results
  for both responses, although different for low $q$ ($q=500$ MeV/c), tend to be 
  very similar for large $q$ (compare the two responses for $q=2$ GeV/c).
\item   Finally, the largest kinematical relativistic effects are observed for
  the two sideways polarized responses, ${\cal R}^{T'}_s$ and ${\cal R}^{TL'}_s$,
  particularly for the latter, where they are typically of the order of $\sim$40--50\%, and even
  much bigger for some specific off-shell prescriptions and large $p$ values.
  This uncertainty is proven to
  diminish (for a given $p$) as $q$ increases. As a conclusion, while
  for very small missing momenta the expansion procedure is not too
  bad, a semi-relativistic treatment is not suitable for these cases
  when intermediate-to-high values of $p$ are considered; indeed, the
  differences from the relativistic results can be already as high as 
  $\sim$25\% for ${\cal R}^{T'}_s$ and $\sim$15\% for ${\cal
  R}^{TL'}_s$ even for missing momenta as low as $p\sim 200$ MeV/c.  
\end{itemize}

Summarizing, we may conclude that the sideways polarized responses are significantly affected by the kinematical relativistic effects. Moreover, once one has selected
the recoil nucleon polarization direction (longitudinal versus sideways)
class `1' responses, ${\cal R}^{TL'}_l$ and ${\cal R}^{T'}_s$, seem to be
less sensitive to relativistic kinematics than the corresponding class `0'
responses, ${\cal R}^{TL'}_s$ and ${\cal R}^{T'}_l$. This agrees with the general 
findings already presented for the unpolarized responses. 
It is important to point out, however, that kinematical relativistic effects seem to be
more sensitive to the recoil nucleon polarization direction than to the 
response class type. Hence, an analysis based on semi-relativistic
reductions should be taken very cautiously in some cases. 

In context, we also note that a similar analysis has been performed in
a parallel study of dynamical relativistic effects~\cite{Cris1} and there a
different pattern emerges. Without presenting any detail here
(referring the reader to that other work) we summarize by stating
that, as a general rule, we find that the
less important the dynamical relativistic effects are found to be, the
larger are the kinematical effects.

To finish the analysis in this section let us recall that the 
longitudinal polarized responses ${\cal R}^{T'}_l$ and ${\cal R}^{TL'}_l$
tend to be similar to the unpolarized ones, ${\cal R}^T$ and ${\cal R}^{TL}$,
respectively, for higher $q$-values. This result can be explained
by noting that as $q$ increases, the range of $\theta_N$ corresponding
to $0\leq p\leq 500$ MeV/c is significantly reduced.
In other words, as $q$ goes up the kinematics become closer to parallel kinematics, i.e.,
the $\theta_N$-range covered within $0\leq p\leq 500$ MeV/c is much
smaller. This means that the higher the momentum $q$, the closer are the directions
of $\nq$ and the longitudinal polarization $\nl$ aligned. In fact, in the limit
case of parallel kinematics ${\cal R}^{TL'}_l$ and ${\cal R}^{TL}$ are zero while
${\cal R}^{T}$ and ${\cal R}^{T'}_l$ coincide.


\subsection{Analysis of single-nucleon responses for parallel kinematics}

\begin{figure}
{\par\centering \resizebox*{0.9\textwidth}{0.44\textheight}
{\includegraphics{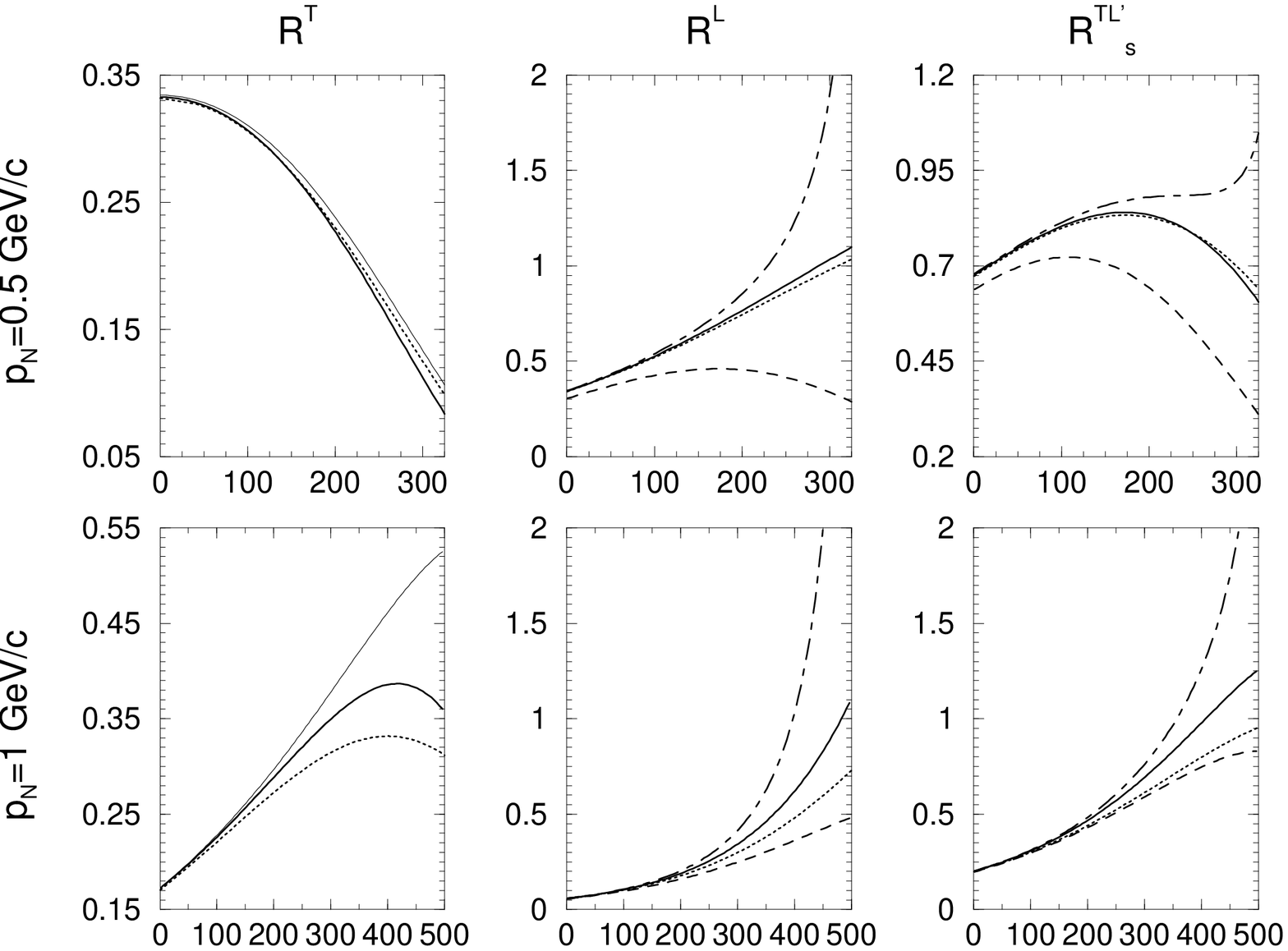}} \par}
\vspace{0.2cm}
{\par\centering \resizebox*{0.9\textwidth}{0.44\textheight}
{\includegraphics{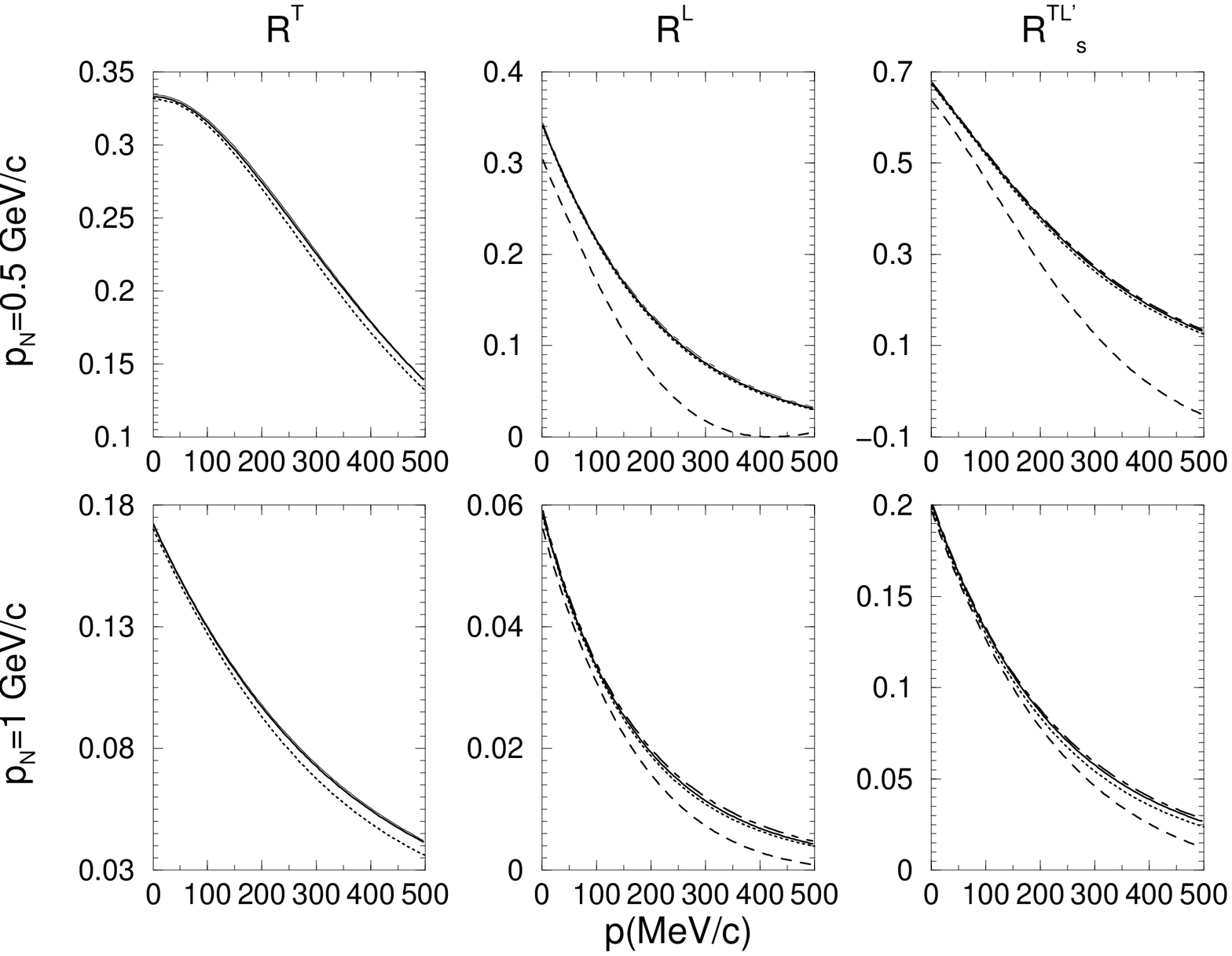}} \par}
\caption{\label{figure16}Single-nucleon responses in parallel kinematics.
Results are shown for two values of the outgoing nucleon momentum. The two top
panels correspond to strictly parallel kinematics, $\theta=0$ (also called the
positive $p$-region), while the two bottom panels correspond to antiparallel kinematics,
$\theta=180^0$ (negative $p$-region). The labels are the same as in Fig.~\ref{figure8}, including the $CC1$ results (thin solid lines) for ${\cal R}^T$.}
\end{figure}

\begin{figure}
{\par\centering \resizebox*{0.9\textwidth}{0.44\textheight}
{\includegraphics{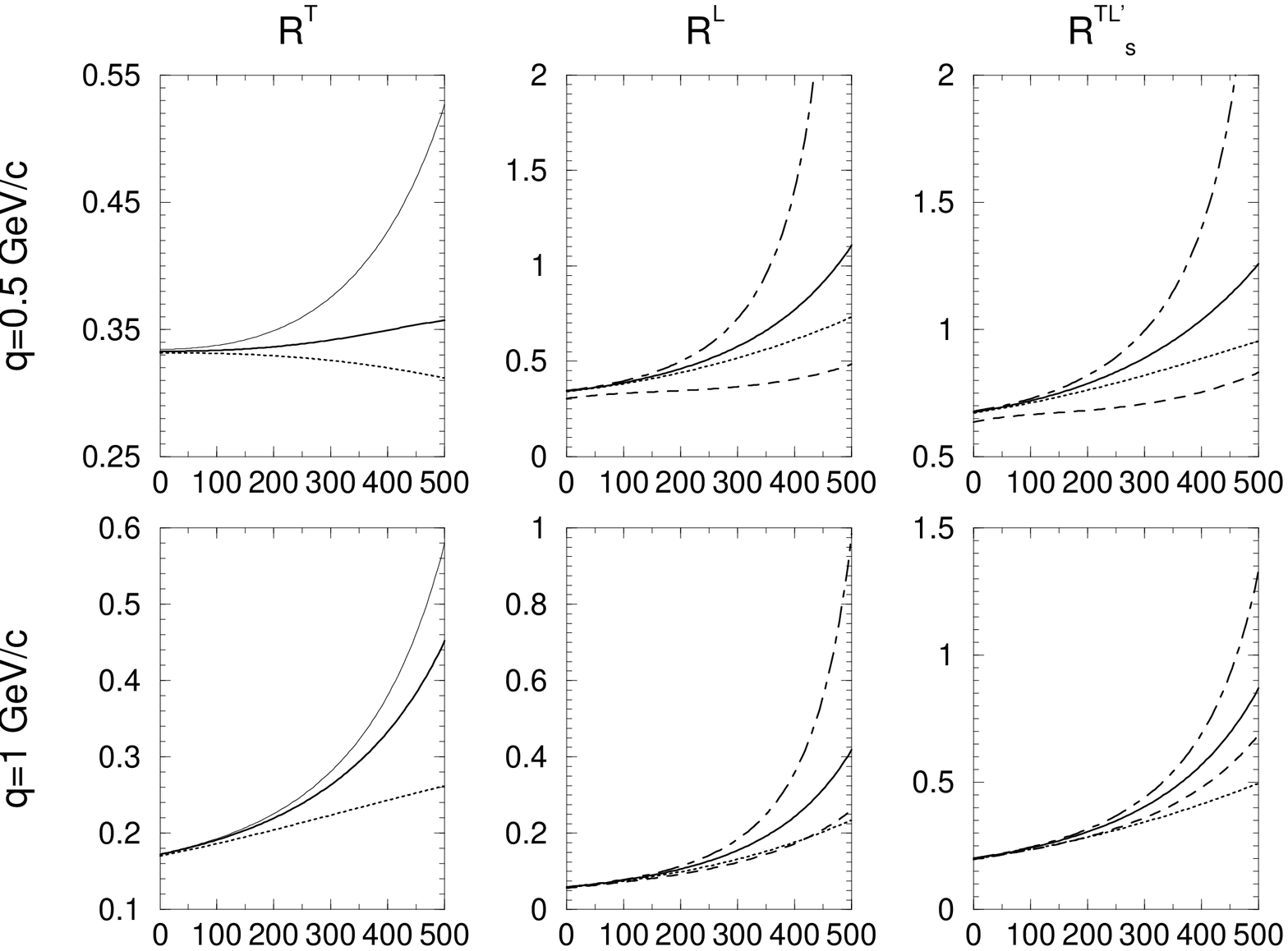}} \par}
\vspace{0.2cm}
{\par\centering \resizebox*{0.9\textwidth}{0.44\textheight}
{\includegraphics{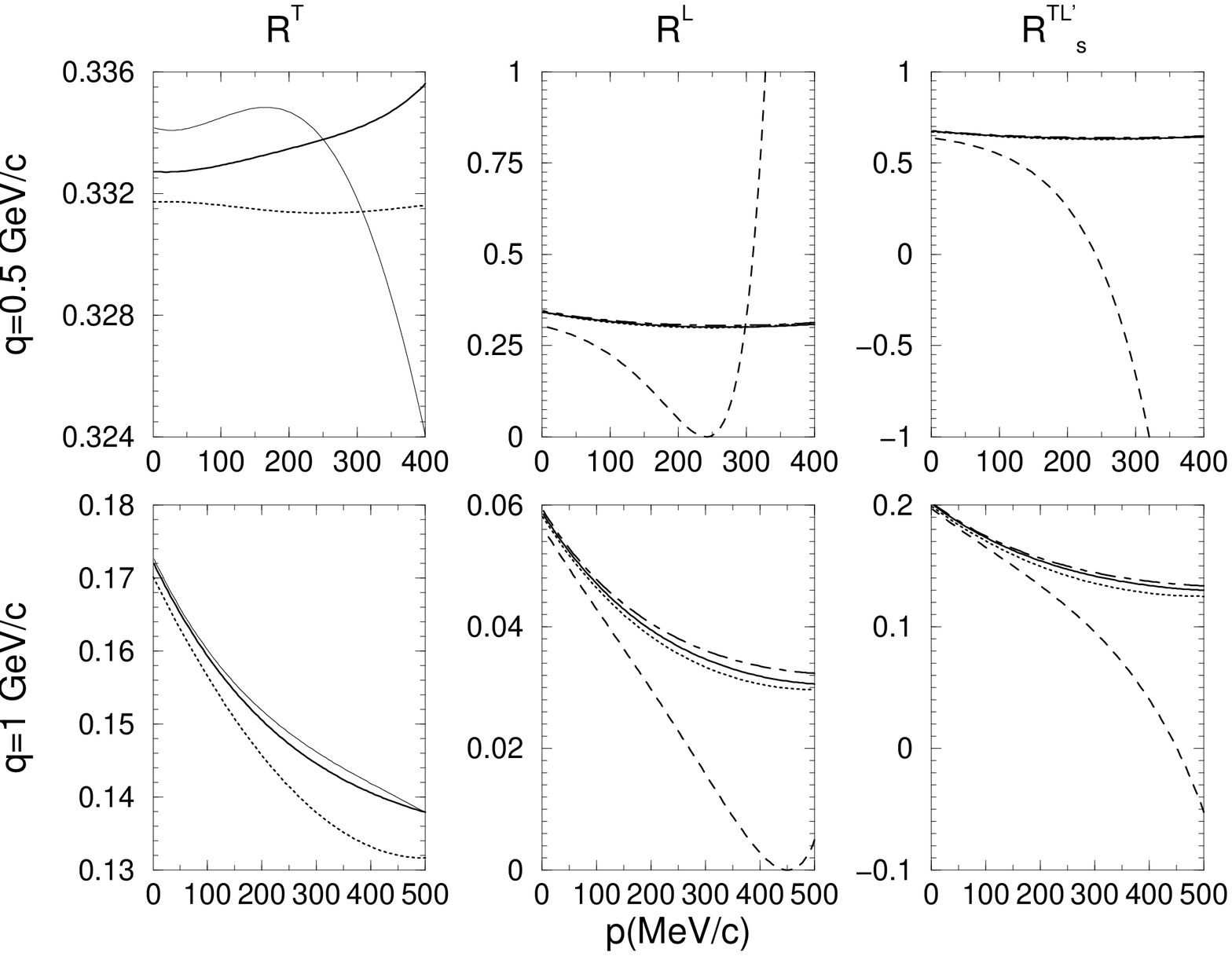}} \par}
\caption{\label{figure17}Same as Fig.~\ref{figure16}, but for fixed $q$-values.}
\end{figure}

In Figs.~\ref{figure16} and \ref{figure17} 
we present the unpolarized and polarized responses that enter in
$A(\vec{e},e'\vec{p})B$ reactions within parallel kinematics and PWIA. Two choices of
the dynamical variables
have been selected: $\{p_N,\theta_N=0,{\cal{E}}=0,p\}$ and
$\{q,\theta_N=0,{\cal{E}}=0,p\}$ (see Section 4 for details). 

Within parallel kinematics and the PWIA only four responses survive in $(\vec{e},e'\vec{N})$ 
processes, ${\cal R}^L$, ${\cal R}^T$, ${\cal R}^{T'}_l$ and 
${\cal R}^{TL'}_s$. Moreover, ${\cal R}^T$ and
${\cal R}^{T'}_l$ are proven to be identical in this case. 
Therefore, in the figures that follow we only present results for
${\cal R}^L$, ${\cal R}^T$ and ${\cal R}^{TL'}_s$. Fig.~\ref{figure16} 
corresponds to the choice
$\{p_N,\theta_N=0\}$, while Fig.~\ref{figure17} corresponds to $\{q,\theta_N=0\}$. In both cases we represent the responses for
$\theta=0$ and $\theta=180^0$. In each graph we show the curves corresponding to
the various off-shell prescriptions based on the $CC2$ current and the `on-shell' approach. 
Again, although not shown in the graphs, we have also explored the behaviour for the $CC1$ prescriptions. 
The semi-relativistic reduction coincides with the $CC1^{(0)}$ prescription in
parallel kinematics, as the expansion has been made in powers of $\chi\equiv (p/M_N)\sin\theta$, 
and this variable is strictly zero for this choice of kinematics.  
The labels are the same as in previous figures.

We start the discussion with the responses obtained by fixing the outgoing
nucleon momentum $p_N$ (Fig.~\ref{figure16}). The results for strictly parallel kinematics 
are displayed in the two
top panels. Here $q<p_N$ and higher $p$-values imply lower values of $q$. 
In this situation there exists a maximum value of $p$ for which $q$ comes
close to $\omega$, i.e. to the real photon point.
In such a case, ${\cal R}^L$ and ${\cal R}^{TL'}_s$ evaluated with the Landau gauge are
divergent because of the term $Q^2$ entering in the denominator in the 
general expressions~(\ref{eq5s4},\ref{eq10s4}). 
When current conservation is imposed (Coulomb and Weyl gauges)
the $Q^2$ factor cancels and the divergency disappears. Nevertheless, in this case using the popular off-shell prescriptions, as we do here, is highly suspect. The situation is clearly
different within antiparallel kinematics (bottom panels).
Here the larger the momentum $p$ is the higher is the momentum transfer $q$.

The results in Fig.~\ref{figure16} show that off-shell uncertainties are clearly enhanced for 
strictly parallel kinematics. In particular, only the Weyl gauge present a different 
behaviour within antiparallel kinematics from the other prescriptions considered.
This difference tends to diminish as $p_N$ grows. On the contrary, the
discrepancy between the various off-shell options explode for large $p$-values
within the strictly parallel kinematics regime. 
Similar comments can be also applied to the `on-shell' prescription:
again the relative differences with the
relativistic responses are much smaller within antiparallel kinematics. It
is interesting to point out that the discrepancy between the `on-shell' approach
and the $CC1^{(0)}$ calculation grows (although less so for negative $p$) as
$p_N$ increases.

In Fig.~\ref{figure17} the responses for the $\{q,y\}$ choice are shown.
The two top panels correspond to $\theta=0$ and two bottom panels
to $\theta=180^0$. The general behaviour of the results for
$\theta=0$ follows similar trends to the one discussed in the previous case. In 
particular, the purely transverse response is very sensitive to the choice of the
current as well as to the `on-shell' approach for higher missing momenta. In the
case of ${\cal R}^L$ and ${\cal R}^{TL'}_s$ the main uncertainty is introduced by
the gauge choice. The exploding behaviour shown by the Landau results for the highest
$p$-values is again explained by the factor $Q^2$ entering in the denominator in
Eqs.~(\ref{eq5s4},\ref{eq10s4}). In this situation ($q$-fixed) as $p$ grows the outgoing nucleon
momentum $p_N$ also increases and consequently the energy transfer $\omega$ comes
higher, approaching the value of the momentum transfer $q$, i.e., the real photon point. For antiparallel kinematics (bottom panels), increasing $p$ implies decreasing $p_N$ and hence $\omega$ being
smaller. This also explains the totally different behaviour shown by the Weyl
results in the longitudinal and transverse-longitudinal responses. Let us recall
that the Weyl gauge implies that ${\cal R}^L$ and ${\cal R}^{TL'}_s$ are evaluated
by taking only the longitudinal (3) components of the single-nucleon tensors (current
conservation is imposed) multiplied by a general factor $(q/\omega)$ or $(q/\omega)^2$.
Thus as $p$ grows the energy transfer $\omega$ goes down and the above factors may give
a very important contribution ($q$-fixed). In particular, for $q=500$ MeV/c the
limit value $p=500$ MeV/c corresponds to $p_N=0$. The energy transfer is then simply
given by $\omega=M_N+E_B-M_A$, which can be approximated by $E_S+\cal{E}$, a very
small value compared with $q$.

Summarizing, we may conclude that within parallel kinematics the ``safest'' situation 
corresponds to $\theta=180^0$ (antiparallel) fixing the kinetic energy of the
outgoing nucleon. Unfortunately, since when compared with strictly parallel kinematics
($\theta=0$) one has higher $|Q^2|$, smaller cross sections and more difficulty performing
measurements under such conditions, fewer data exist for antiparallel kinematics. In the 
antiparallel case, apart from the Weyl gauge choice that maximizes
the discrepancy, the results for the remaining off-shell and
`on-shell' prescriptions are all rather similar. The spread of the 
various approaches gets wider for antiparallel kinematics and $q$-fixed. Particularly, the
Weyl gauge tends to explode for high $p$ in the $L$ and $TL'$ responses. Finally, the
behaviour of the responses for strictly parallel kinematics is 
rather similar in the two situations analyzed, $p_N$ and $q$ fixed. In both cases,
the ambiguities introduced by the different approaches tend to be very high for
large $p$-values. Moreover, as $p$ reaches very high values one is
moving close to the real photon point. In that case the Landau gauge responses diverge.


\subsection{Transferred polarization asymmetries}


In this section we analyze the transferred polarization asymmetries introduced
in Eq.~(\ref{eq11s3}). These observables are given as ratios between polarized
and unpolarized responses, where one hopes to gain different insight
into the underlying physics from what may be revealed through the
responses themselves. For instance, it may (or may not) be true that
the polarization transfer asymmetries are less affected by FSI
effects, by off-shell ambiguities, etc., than are the responses. 
One goal of this work is to explore briefly a few of these issues. In particular, there exists considerable interest in 
measurements of $P'_l$ and $P'_s$ as they may provide information on
the structure of the nucleon in the nuclear medium. A more complete analysis
of the off-shell and dynamical relativistic effects in these
observables is presented in~\cite{Cris1}, and hence
here we focus mainly on the kinematical relativistic effects within PWIA. 

In Figs.~\ref{figure18} and \ref{figure19} 
we show results for $P'_l$ and $P'_s$, respectively. Let us recall that
the responses in PWIA factorize. This means that the transferred polarizations do not
depend on the nuclear dynamics, and only the single-nucleon responses contribute. As is well-known,
this constitutes a basic difference with respect to the relativistic plane-wave impulse
approximation (RPWIA)~\cite{Cris1,Cab98a}.
To simplify the discussion that follows we only present results
for $\{q,y\}$-fixed kinematics. The excitation energy $\cal{E}$ and the azimuthal
angle $\phi$ are taken to be zero. The electron scattering angle selected is $\theta_e=30^0$,
i.e., forward electron scattering. This kinematical
situation maximizes the off-shell effects~\cite{Cris1}. The labels of the curves
in each graph are the same as those used for the single-nucleon responses. 

We may summarize the basic findings as follows:

\begin{figure}
{\par\centering \resizebox*{1.0\textwidth}{0.75\textheight}
{\includegraphics{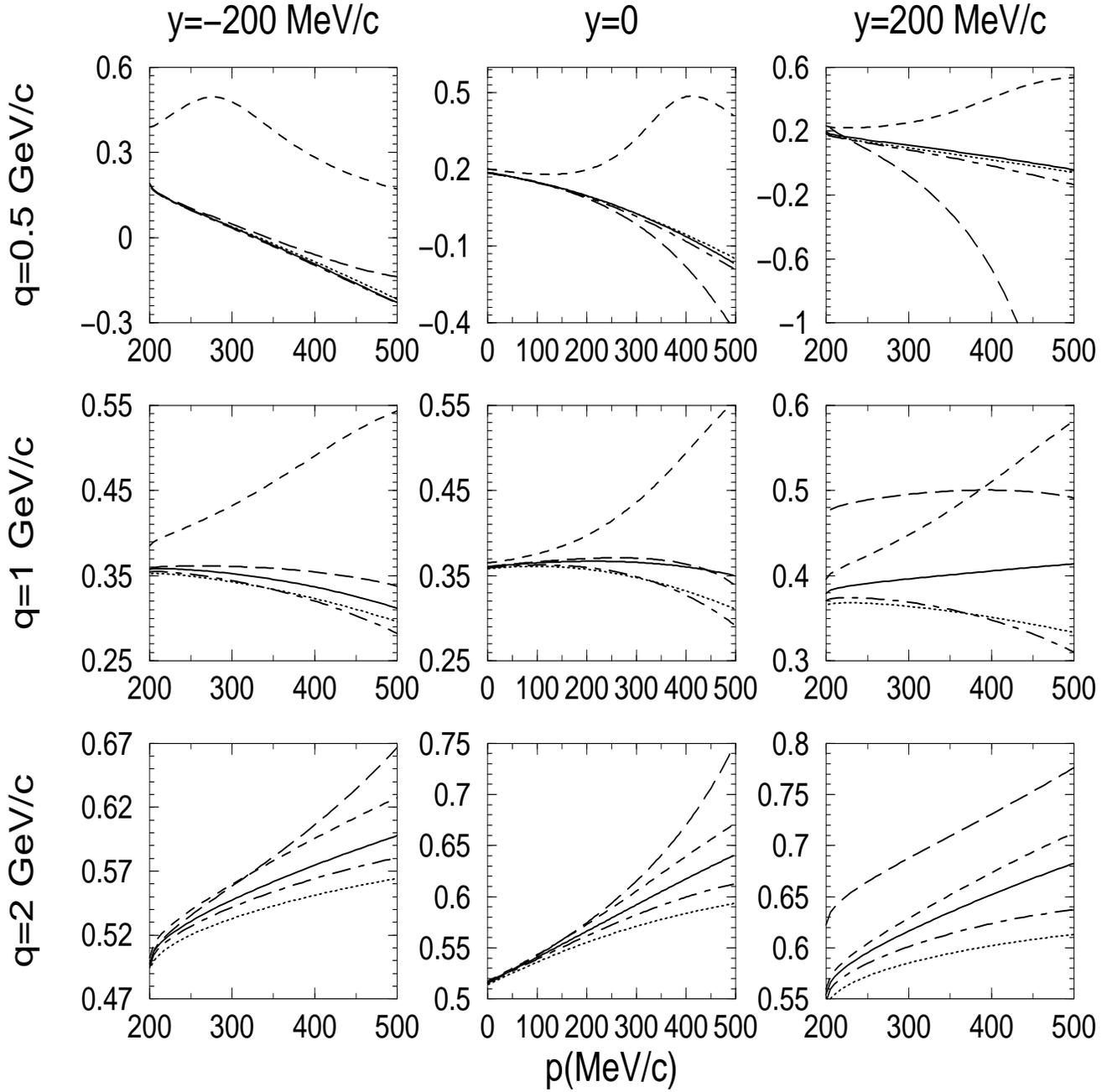}} \par}
\caption{\label{figure18}Transferred polarization asymmetry for longitudinal spin
direction. The labels are as in Fig.~\ref{figure8}.}
\end{figure}

\begin{figure}
{\par\centering \resizebox*{1.0\textwidth}{0.75\textheight}
{\includegraphics{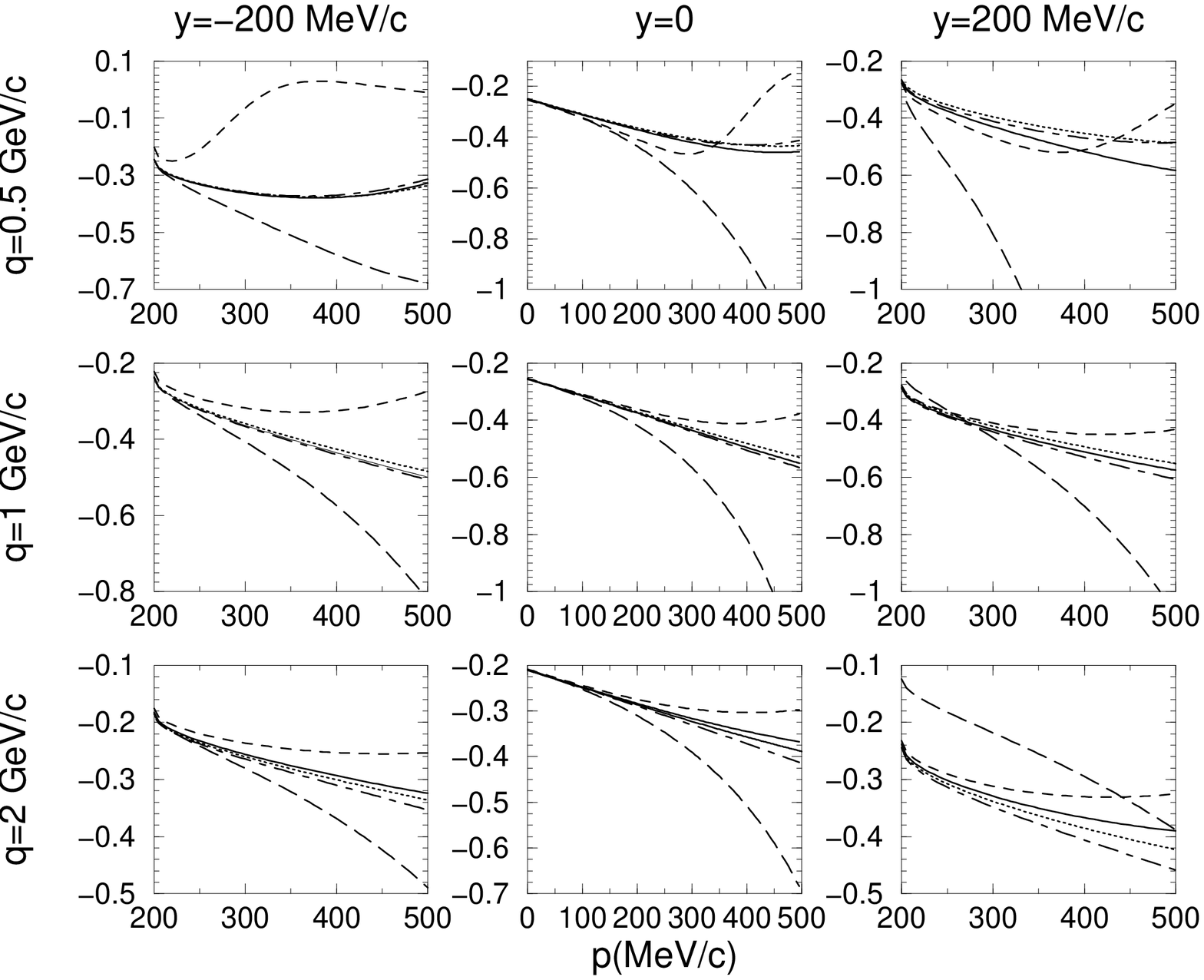}} \par}
\caption{\label{figure19}Same as Fig.~\ref{figure18}, but for the transferred polarization 
asymmetry for the sideways spin direction.}
\end{figure}

\begin{itemize}
\item
      The largest off-shell effects are introduced by the Weyl gauge (short-dashed lines). 
      This is consistent with the results
      already observed for the single-nucleon responses. Moreover, the discrepancy
      between the Weyl results and the Landau or Coulomb responses tends to
      diminish for higher $q$. In contrast, the
      discrepancies between the remaining off-shell prescriptions are rather
      small for low $q$ and negative $y$.
\item
      Concerning the `on-shell' approach (dotted line), the results
      obtained are rather similar
      to those for either the Coulomb or Landau gauges. Actually, they lie amongst the 
$CCi^0$ and $NCCi$ ($i=1,2$) results except for $P'_l$ and $q=2$ MeV/c; even in that 
case the off-shell uncertainty introduced by adding the `on-shell' result remains quite small, 
as one can easily appreciate from the figures. 
\item
      The results obtained for the semi-relativistic reduction (long-dashed line) show
      that the kinematical relativistic effects may have a very
      important role to play, modifying
      completely the structure of the polarization asymmetries $P'_l$
      and $P'_s$ for some kinematics. To be more specific, the
      semi-relativistic reductions compare less well with the relativistic results for 
      the sideways polarization, as was expected from the fact that the sideways polarized 
      responses were the most sensitive to kinematical relativistic effects. The only situations 
      for which a semi-relativistic reduction may work reasonably well are the followings:
\begin{itemize}
\item
  The low-$p$ region ($p\leq150$ MeV) when we are under quasielastic
  conditions where $y=0$: the differences between relativistic and
  semi-relativistic results remain small (less than 10\%) for all
  kinematics except for the case 
  $q=500$ MeV/c for $P'_l$. Going to higher missing 
  momenta implies finding larger kinematical effects 
  (bigger than 10\%) except for the $q=1,2$ MeV/c cases 
  in $P'_l$, for which the semi-relativistic reduction 
  approximates the relativistic results rather well 
  (within 6\% up to $p=350$ MeV/c).
\item
  For the $y=-200$ MeV/c cases, the kinematical relativistic effects are small up 
  to $p=250$ MeV/c in the case of $P'_s$ and up to $p=350$ MeV/c for $P'_l$ (again 
  except for $q=500$ MeV/c). However, for $y=200$ MeV/c, the semi-relativistic reductions 
  differ significantly from the relativistic results for both observables. 
\end{itemize}
Consequently, one has to be very careful before using a
semi-relativistic reduction for these observables, given that only
in particular regimes does it provide results that are similar to the relativistic ones. 
Even for relatively small $p$-values one can find rather important kinematical relativistic 
effects. Finally, it is interesting to point out that the semi-relativistic approach may 
even yield non-physical results, transfer polarizations smaller than -1.
\end{itemize}

Summarizing, we may conclude in general that the transferred polarization asymmetries evaluated with
the Landau and Coulomb prescriptions do not show large differences; 
they are at most
of the order of $\sim$1.5\% for $p\approx 100$ MeV/c, where the cross section and
response functions are mainly located in the case of $^{16}$O. The Weyl gauge 
on the contrary gives rise to
results that sometimes differ completely. This is in accordance with the results discussed
for the responses in the previous section and it also agrees with several 
previous papers~\cite{Cris1,Cab98a,Cab93}. The `on-shell' approach in general fits rather nicely
with the off-shell results based on either Landau or Coulomb gauges.

Taking into account the previous comments on semi-relativistic reductions, 
it turns out that even in approximation schemes where the momentum and energy transfers are treated
exactly it does not seem to be very appropiate to employ such analyses of
the transferred polarization observables for the
$\{q,y\}$-fixed kinematics, except perhaps at rather low missing
momenta. 


\section{Summary and conclusions}


In this work our interest has been focused on the analysis of kinematical
relativistic effects and off-shell uncertainties in $A(\vec{e},e'\vec{N})B$
reactions within the context of the plane-wave impulse
approximation (PWIA). In other work being done in parallel we have
also been exploring dynamical relativistic effects using 
the relativistic plane-wave impulse approximation (RPWIA)~\cite{Cris1}.
The basic difference between RPWIA and PWIA is that the latter
does not include the role played by the enhancement of the negative-energy components
of the bound nucleon wave function. A crucial consequence of this is that within PWIA the
response functions (differential cross section) factorize into single-nucleon
responses (single-nucleon cross section) and the spectral function or momentum
distribution, the latter containing the entire dependence on the nuclear dynamics. 
Therefore, observables given as ratios of responses or cross sections, such as the transferred
polarization asymmetries, do not depend on the nuclear dynamics in PWIA.

In this work we have presented a systematic analysis for a variety of
kinematical conditions of all of the single-nucleon responses that
enter in the description of $A(\vec{e},e'\vec{N})B$ processes within
PWIA. In particular, we have analyzed the results obtained
not only for the quasielastic peak conditions, but also moving far
above or below the peak. In each kinematical situation
a very wide set of different options has been considered:
\begin{itemize}
\item First, concerning fully-relativistic PWIA calculations: here various alternatives to deal with
the off-shell character of the bound nucleon have been explored. They are connected with the
current operator choice and the current conservation property. Additionally, within this
scheme, we have also analyzed the so-called `on-shell' prescription in which the bound
nucleon is forced to be on-shell. 
\item Second, we have also presented a detailed study of the
reductions of the various responses
treating exactly the problem of the energy and
momentum transfers, but expanding in initial state momenta divided by the nucleon
mass -- indeed, these are not the traditional non-relativistic expansion schemes
and we have highlighted this fact by calling them ``semi-relativistic'' expansions.
Different semi-relativistic alternatives have been explored.
This analysis gives us a clear image of the role played by the kinematical
relativistic effects in the various observables that are accesible in
$(\vec{e},e'\vec{N})$ reactions. Moreover, the semi-relativistic reduction has
allowed us to classify the responses into three basic categories according to
the leading order term in the expansion: class ``0''
(leading-order $\chi^0$), class ``1'' ($\chi^1$) and class ``2'' ($\chi^2$) responses.
We have found this to be a very natural way to classify the responses and,
even more importantly, it allows us to simplify the discussion of the role
of the kinematical and dynamical relativistic effects.
\end{itemize}

Concerning the results for $\{q,y\}$-fixed kinematics, we may summarize the basic
conclusions as follows. The largest off-shell uncertainties are introduced in general
by the prescriptions 
based on the Weyl gauge. This agrees with the general discussion already presented
in~\cite{Cris1,Cab98a}. Furthermore, we also find that the discrepancy between the
Weyl results and the Landau and Coulomb gauge results becomes smaller for higher values of the
momentum transfer $q$ and increasing values of the scaling variable $y$. With regard to
the results evaluated with the Landau and Coulomb gauges, the discrepancy between them
is much smaller, increasing slightly as $y$ goes from negative to positive values.
Finally, the ambiguity introduced by the choice of the current operator
is also well under control. The `on-shell' approach gives rise
to results which are rather similar to the off-shell ones based on the Coulomb and
Landau gauges for $q$ and $y$ small. We have revealed the reason for this discrepancy 
and have explained why it increases for higher $q$ and $y$-values. 

Kinematical relativistic effects are clearly made evident when comparing the semi-relativistic
calculation with the various off-shell results. Here we find as a general rule that
the discrepancy between the two approaches is significantly smaller for higher
class type responses, i.e., kinematical relativistic effects are very small for class
``2'' responses, increase slightly for class ``1'' responses, and are 
the largest for class ``0''
responses. This behaviour is strictly true in the case of the four unpolarized responses:
${\cal R}^L$, ${\cal R}^T$ (class ``0''), ${\cal R}^{TL}$ (class ``1'') and
${\cal R}^{TT}$ (class ``2''). For the recoil nucleon polarized responses, the direction
of the spin polarization plays also a crucial role in the determination of the relativistic
effects, and one should take into account both ingredients, polarization direction and
response class type. We have found that kinematical relativity shows a stronger sensitivity
to the recoil nucleon polarization direction than to the response class type. However,
it is important to point out that for a fixed polarization direction, the above statement
on the connection between kinematical relativistic effects and class type responses
remains valid.

In this work we have also presented a careful analysis of the results corresponding to
parallel kinematics. Here one should distinguish between strictly parallel kinematics, i.e.,
${\np}$, $\nq$ and $\np_N$ parallel, and antiparallel kinematics, i.e., $\np$ opposite
to $\nq$ and $\np_N$. Two different choices for the dynamical variables that specify
the response functions have been considered. In the former the outgoing nucleon momentum $p_N$
is fixed, and in the latter the momentum transfer $q$ is assumed fixed. The results
show very clearly that the situation corresponding to antiparallel kinematics and 
$p_N$-fixed minimizes the off-shell uncertainties. On the contrary, the three remaining regimes, strictly parallel kinematics with
$p_N$ fixed and the two kinematics corresponding to $q$-fixed, present large uncertainties.
In particular, within strictly parallel kinematics, as the missing momentum $p$ increases one
is approaching the real photon point for which the difference
between the various prescriptions can explode --- in fact, the Landau results diverge in
that case.

The transferred polarization asymmetries, $P'_l$ and $P'_s$, may provide information
on nucleon structure in the nuclear medium, and accordingly these 
have been also analyzed in the case of $\{q,y\}$-fixed kinematics
and forward electron scattering. As is well-known, this situation maximizes the off-shell 
uncertainties. We find that the results for the off-shell prescriptions based on the
Landau and Coulomb gauges and the `on-shell' approach present a similar behaviour whose
relative differences depend on the specifics of the kinematics. The Weyl
prescription produces results that are completely different, particularly for low-$q$
and negative-$y$. This discrepancy tends to diminish for higher $q$. Finally, the
kinematical relativistic effects are also shown to be very important, in particular
for the sideways nucleon polarization direction. In this case, the semi-relativistic
results present a behaviour that differs significantly from all other calculations
independently of the specific kinematics selected. The semi-relativistic approach may
even lead to unphysical results $|P'_s|>1$. Although the sensitivity of $P'_l$ to
kinematical relativistic effects is reduced, one still observes very important
deviations even at the quasielastic peak.

Although FSI and other ingredients such as two-body meson exchange-currents may
have additional important effects in the analysis of $(\vec{e},e'\vec{N})$ observables
(work along this line is presently in progress), some of the results presented 
here within PWIA are by themselves very illustrative of the uncertainties introduced by
different approximations. In particular, the results for the transferred polarization
asymmetries tell us that semi-relativistic 
treatments may not be adequate when describing such
observables even for low-$q$ and quasielastic peak conditions. Hence great caution
should be taken when distorted-wave impulse approximation (DWIA) calculations 
based on semi-relativistic --- or, even more so, non-relativistic --- reductions of the current are used to describe experimental transferred polarization asymmetries.

\subsection*{Acknowledgements}
This work was partially supported by funds provided by DGICYT (Spain) 
under Contracts Nos. PB/98-1111, PB/98-0676 and the Junta
de Andaluc\'{\i}a (Spain) and in part by the U.S. Department of Energy under
Cooperative Research Agreement No. DE-FC02-94ER40818.
M.C.M acknowledges support from a fellowship from the Fundaci\'on
C\'amara (University of Sevilla). J.A.C. also acknowledges financial support from MEC (Spain) for a sabbatical stay at MIT (PR2001-0185). The authors thank E. Moya, J.M. Ud\'{\i}as and J.R. Vignote for
their helpful comments.


\appendix


\section*{Appendix A}


In this appendix we present the explicit expressions for the symmetric and antisymmetric
terms of the single-nucleon tensor ${\cal W}^{\mu\nu}$ that enters in the analysis of
$A(\vec{e},e'\vec{N})B$ reactions withing PWIA for both
current operators, $CC1$ and $CC2$, as well as the analytic expressions for the polarized single-nucleon $CC1^{(0)}$ and $CC2^{(0)}$ responses. The tensors are given as
\begin{itemize}

\item {\bf $CC1$ current}

\ba
{\cal S}^{\mu\nu}&=& \frac{1}{2M^{2}_{N}}\left\{(F_{1}+F_{2})^{2}
\left(\overline{P}^{\mu}P^{\nu}_{N}+\overline{P}^{\nu}P^{\mu}_{N}+
\frac{\overline{Q}^{2}}{2}g^{\mu\nu}\right) \right. \nonumber \\
&-& \left. \left[(F_{1}+F_{2})F_{2}
-\frac{F_{2}^{2}}{2}\left(1-\frac{\overline{Q}^{2}}{4M^{2}_{N}}\right)\right]
C^{\mu}C^{\nu}\right\}
\ea

\ba
{\cal A}^{\mu\nu} &=& \frac{i}{2M^{2}_{N}}\left\{M_{N}^{2}(F_{1}+F_{2})^{2}
     \varepsilon^{\alpha\beta\mu\nu}S_{N\alpha}\overline{Q}_\beta \right. \nonumber \\
      &+&\left. (F_{1}+F_{2})\frac{F_{2}}{2M_{N}}\left(C^{\mu}
      \varepsilon^{\alpha\beta\gamma\nu}-C^{\nu}
      \varepsilon^{\alpha\beta\gamma\mu}\right)\overline{P}_{\alpha}P_{N\beta}
       S_{N\gamma}\right\}
\ea

\item {\bf $CC2$ current}

\ba
& & {\cal S}^{\mu\nu}=\frac{1}{2M_{N}^{2}}\left\{ F^{2}_{1}
\left(\overline{P}^{\mu}P^{\nu}_{N}+\overline{P}^{\nu}P^{\mu}_{N}+
\frac{\overline{Q}^{2}}{2}g^{\mu\nu}\right)+F_{1}F_{2}\left(
Q\cdot\overline{Q}g^{\mu\nu}-\frac{\overline{Q}^{\mu}Q^{\nu}
+\overline{Q}^{\nu}Q^{\mu}}{2}\right)\right. \nonumber \\
&+& \left.
\frac{F_{2}^{2}}{4M^{2}_{N}}\left[P_{N}\cdot Q\left(\overline{P}^{\mu}Q^{\nu}
+\overline{P}^{\nu}Q^{\mu}\right)+\overline{P}\cdot Q
\left(P^{\mu}_{N}Q^{\nu}+P^{\nu}_{N}Q^{\mu}\right)-Q^{2}\left(
P^{\mu}_{N}\overline{P}^{\nu}+P^{\nu}_{N}\overline{P}^{\mu }\right) \right. \right. \nonumber
\\
&-& \left. \left.
\left(2M^{2}_{N}-\frac{\overline{Q}^{2}}{2}\right)
Q^{\mu}Q^{\nu}+g^{\mu\nu}\left(2M^{2}_{N}Q^{2}-\frac{Q^{2}
\overline{Q}^{2}}{2}-2P_{N}\cdot Q\overline{P}\cdot Q\right)\right] \right\}
\ea

\ba
& & {\cal A}^{\mu\nu}=\frac{i}{2M^{2}_{N}}\left\{
-M_{N}F^{2}_{1}\varepsilon^{\alpha\beta\mu\nu}\overline{Q}_{\alpha}
S_{N\beta}\right. \nonumber \\
&+& \left. 
\frac{F_{1}F_{2}}{M_{N}}\left[\varepsilon^{\alpha\beta\mu\nu}
(\overline{P}\cdot QP_{N\alpha}S_{N\beta}-M^{2}_{N}Q_{\alpha}S_{N\beta})
+\frac{1}{2}\left(Q^{\mu}\varepsilon^{\alpha\beta\gamma\nu}-
Q^{\nu}\varepsilon^{\alpha\beta\gamma\mu}\right)
S_{N\alpha}P_{N\beta}\overline{P}_{\gamma}\right]
\right. \nonumber \\
&-& \left.
\frac{F^{2}_{2}}{4M_{N}}\left[
\varepsilon^{\alpha\beta\mu\nu}(2\overline{P}\cdot QS_{N\alpha}
Q_{\beta}+Q^{2}C_{\alpha}S_{N_{\beta}})+ 
(Q^{\mu}\varepsilon^{\alpha\beta\gamma\nu}-Q^{\nu}\varepsilon^{\alpha\beta\gamma\mu})
S_{N\alpha}C_{\beta}Q_{\gamma}\right]\right\} \, , \nonumber
\\
& & 
\ea
where we have defined $C^{\mu}\equiv(\overline{P}+P_{N})^{\mu}$.

\begin{table}
{\centering \begin{tabular}{ccccc}
\hline
Axis & \( \mu=0 \) & \( \mu=1 \)& \( \mu=2 \) & \( \mu=3 \)
\\
\hline
\\
$\nl$ & $\sqrt{\gamma_N^2-1}$ & $\frac{\gamma_N\chi}{\sqrt{\gamma_N^2-1}}$ 
    & 0 & $\frac{\gamma_N(\chi'+2\kappa)}{\sqrt{\gamma_N^2-1}}$ 
\\
\\
$\ns$ & 0 & $\frac{\chi'+2\kappa}{\sqrt{\gamma_N^2-1}}$ & 0 & $-\frac{\chi}{\sqrt{\gamma_N^2-1}}$ 
\\
\\
$\nn$ & 0 & 0 & 1 & 0 \\
\hline 
\end{tabular}\par}
\caption{\label{tab2} Components of the spin four-vector 
$S_N^\mu(\nl,\ns,\nn)$ in the laboratory
system (see text for the notation).}
\end{table}

By using the expressions given in Eqs.~(\ref{eq9s4},\ref{eq10s4}) and referring the recoil nucleon spin four-vector $S_N^\mu$ to
the coordinate system defined by the axes $\nl$, $\ns$ and $\nn$ (see Table 1), it is possible to write down 
in a very compact form specific answers for the
polarized single-nucleon $CC1^{(0)}$ and $CC2^{(0)}$
responses. They are given as
\ba
& &{\cal R}^{T'}_l = \frac{2}{\sqrt{\gamma_N^2-1}}\left\{
(F_1+F_2)\left[\frac{\xi\overline{\tau}}{\kappa}
  \left(\gamma_NF_1+\overline{\lambda}F_2\sqrt{\frac{\rho_1+\rho_2}{2}}\right)
-\kappa\left(F_1-\overline{\tau}\sqrt{\frac{\rho_1+\rho_2}{2}}F_2 \right)\right]
\right. \nonumber \\
&-& \left.
\frac{\overline{\tau}}{\kappa}\sqrt{\frac{\rho_2-\rho_1}{2}}F_2
\left[\kappa\gamma_N\left(F_1-\overline{\tau}\sqrt{\frac{\rho_1+\rho_2}{2}}F_2\right)
-\frac{\xi}{\kappa}\overline{\tau}\left(F_1-(\xi\gamma_N-1)F_2
        \sqrt{\frac{\rho_1+\rho_2}{2}}\right)\right]\right\} \nonumber \\
& &
\label{rtprimel}
\ea
\ba
{\cal R}^{T'}_s &=& \frac{2\chi}{\sqrt{\gamma_N^2-1}}\left\{
    (F_1+F_2)\left[\overline{\lambda}F_1-\gamma_N\overline{\tau}F_2
          \sqrt{\frac{\rho_1+\rho_2}{2}}\right] \right. \nonumber \\
&-& \left.
\frac{\overline{\tau}}{\kappa}\sqrt{\frac{\rho_2-\rho_1}{2}}F_2
\left[F_1(\xi\gamma_N-1)+\overline{\tau}F_2\sqrt{\frac{\rho_1+\rho_2}{2}}
\right]\right\} 
\label{rtprimes}
\ea
\ba
{\cal R}^{TL'}_l &=& \frac{2\sqrt{2}\kappa\chi}{\sqrt{\gamma_N^2-1}}\left\{ 
    (F_1+F_2)\left[(\gamma_NF_1+\overline{\lambda}F_2)
   +\gamma_N\overline{\tau}F_2\left(1-\sqrt{\frac{\rho_1+\rho_2}{2}}\right)\right]
\right. \nonumber \\
&+& \left. 
\frac{\overline{\tau}}{\kappa}\sqrt{\frac{\rho_2-\rho_1}{2}}F_2
    \left[(1-\overline{\lambda}\gamma_N)F_1-(\gamma_N^2-1)F_2\right]\right\}
           \cos\phi \nonumber \\
&&\label{rtlprimel} \\
{\cal R}^{TL'}_s &=& \frac{2\sqrt{2}}{\sqrt{\gamma_N^2-1}}\left\{ 
   (F_1+F_2)\left[\xi\left(\overline{\lambda}F_1-\gamma_N\overline{\tau}F_2
    \sqrt{\frac{\rho_1+\rho_2}{2}}\right)+\kappa^2(F_1+F_2)\right] \right. \nonumber \\
&-& \left.
\frac{\overline{\tau}}{\kappa}F_2\sqrt{\frac{\rho_2-\rho_1}{2}}
    \left[\left(\xi(\gamma_N^2-1)-\gamma_N\kappa^2\right)F_1+
    \overline{\lambda}(\xi^2-\kappa^2)F_2\right]\right\}
    \cos\phi \nonumber \\
&& \label{rtlprimes} \\
{\cal R}^{TL'}_n &=& -2\sqrt{2}\left\{
    \kappa(F_1+F_2)\left[F_1-\overline{\tau}F_2\sqrt{\frac{\rho_1+\rho_2}{2}}\right]
-\frac{\xi^2}{\kappa^2}\overline{\tau}\overline{\lambda}\sqrt{\frac{\rho_2-\rho_1}{2}}F_1F_2
\right\}\sin\phi \, ,\nonumber \\
& & \label{rtlprimen}
\ea
where  we have introduced the usual dimensionless
variables: $\chi\equiv\frac{p}{M_N}\sin\theta$, $\kappa\equiv\frac{q}{2M_N}$,
$\overline{\lambda}\equiv\frac{\overline{\omega}}{2M_N}$,
$\gamma_N\equiv\frac{E_N}{M_N}$,
$\xi\equiv\frac{E_N+\overline{E}}{2M_N}$ and
$\overline{\tau}\equiv \frac{|\overline{Q}^2|}{4M_N^2}$. More details on these 
variables and the relations held between them are given in~\cite{Cab93}. For compactness the above expressions have been written introducing the parameters
$\rho_1$ and $\rho_2$ (see~\cite{Cab93} for details). For the $CC1^{(0)}$
prescription one has $\rho_1=\rho_2=1$, whereas for $CC2^{(0)}$ one has
$\rho_1=\tau/\overline{\tau}$ and
$\rho_2=2[1+\overline{\lambda}(\overline{\lambda}-\lambda)/\tau]^2-\rho_1$
with $\lambda\equiv \frac{\omega}{2M_N}$ and $\tau\equiv \frac{|Q^2|}{4M_N^2}$.
Note that for the $CC1^{(0)}$ prescription one has $\rho_2-\rho_1=0$,
and consequently the single-nucleon responses simplify considerably.
\end{itemize}






\end{document}